# Polarization Measurements of Arecibo–Sky Pulsars: Faraday Rotations and Emission-Beam Analyses

Joanna Rankin[1★], Arun Venkataraman[2], Joel M. Weisberg[3,4], Alice P. Curtin[3,5,6]
[1]Physics Department, University of Vermont, Burlington, VT 05405, USA
[2]Arecibo Observatory, bo. La Esperanza, P.P. Box 53995, Arecibo, Puerto Rico, 0612
[3]Department of Physics and Astronomy, Carleton College, Northfield, MN USA 55057
[4]Orcid ID https://orcid.org/0000-0001-9096-6543
[5]Department of Physics, McGill University, Montreal QC H3A 2T8, Canada
[6]Trottier Space Institute, McGill University, Montreal, QC H3A 2A7, Canada

## ABSTRACT

We present Faraday Rotation Measure (RM) values derived at L- and P-band as well as some 60 Stokes-parameter profiles, both determined from our longstanding Arecibo dual-frequency pulsar polarimetry programs. Many of the RM measurements were carried out toward the inner Galaxy and the Anticenter on pulsars with no previous determination, while others are re-measurements intended to confirm or improve the accuracy of existing values. Stokes-parameter profiles are displayed for the 58 pulsars for which no meaningful Stokes profile at lower frequency is available and four without a high frequency pair. This is a population that includes many distant pulsars in the inner Galaxy. A number of these polarized pulse profiles exhibit clear interstellar-scattering tails; nonetheless, we have attempted to interpret the associated emission-beam structures and to provide morphological classifications and geometrical models where possible.

## 1 INTRODUCTION

Polarimetric observations of pulsars have been conducted with the Arecibo telescope in the 21-cm band since the mid-1980s (*e.g.,* Stinebring et al., 1984). One of the challenges of radio polarimetry is the fact that linearly polarized radio waves propagating through the magnetized interstellar medium (ISM) suffer a wavelength-dependent position-angle rotation, which must be compensated for in order to observe a source over any significant bandwidth. This phenomenon, known as Faraday rotation, can in turn be used to study the magnetized interstellar medium (hereafter, ISM) along the path between a pulsar and the Earth.

Pulsars are usually highly polarized sources whose signals thus exhibit significant Faraday rotation. Therefore, the magnitude of the rotation must be accurately measured in order to compensate for the polarized position angle (PPA) rotation across the passband in order produce sensitive full-Stokes-parameter pulsar profiles, such as those we present later in the Appendix.

The Faraday rotation $\chi$ (in radians) of the linearly polarized position angle $\chi$ at a given wavelength $\lambda$ is given by

$$\chi = RM\ \lambda^2, \qquad (1)$$

where *RM* is the rotation measure in units of rad-m$^{-2}$ and $\lambda$ is the wavelength in m. While the total rotation $\chi$ is not directly measurable because the original PPA at the source is inaccessible to the observer, the *change* in $\chi$ across the observing band is readily measurable for all pulsars across any practical bandwidth at $\lambda \sim 20$ cm, and $\chi$ itself can then be extracted from this information, as detailed below in Section 3.

The rotation measure in turn is the path integral along the source-observer sightline of the electron-density weighted by the longitudinal component of the magnetic field,

$$\mathrm{RM}\ \ (\mathrm{rad\ m^{-2}}) = \frac{\mathrm{e}^3}{2\pi \mathrm{m_e^2 c^4}} \int_{\mathrm{source}}^{\oplus} \mathrm{n_e} \mathbf{B} \cdot \mathrm{d}\mathbf{s} = \int_{\mathrm{source}}^{\oplus} \mathrm{n_e} \mathbf{B} \cdot \mathrm{d}\mathbf{s}/1.232, \qquad (2)$$

where e and $\mathrm{m_e}$ are the charge magnitude and mass of the electron, and c is the speed of light. The constant in the last expression requires electron density $\mathrm{n_e}$ in cm$^{-3}$, magnetic field $\mathbf{B}$ in $\mu$G, and path element d$\mathbf{s}$ in pc. Thus, in measuring the rotation measure for a number of pulsars, one can study the magnetic fields along their lines of sight.

In this paper, we report *RM* values and their uncertainties derived from L- and P-band measurements acquired in Arecibo pulsar polarimetry programs conducted from early 2003, until just prior to the catastrophic collapse of the telescope on 1 December 2020. Several of our later observing programs focused specifically on acquiring a more complete sample of Arecibo pulsar RMs at 20 cm wavelength. Rotation measure values were already available for some of the pulsars in the Arecibo sky, principally as measured in the previous Arecibo RM survey (Weisberg et al., 2004).

The results below reflect a combination of our new measurements on weaker Arecibo pulsars and remeasurements of brighter ones; the latter generally having smaller uncertainties resulting from the great sensitivity of the upgraded Arecibo telescope, receivers, and back ends. These and other pulsar RMs will be used to study the Galactic magnetic field in a companion paper (Curtin, Weisberg & Rankin 2023, *in prep*).

★ E-mail: Joanna.Rankin@uvm.edu

These observing programs also yielded full-Stokes-parameter pulsar profiles. Most of the observing programs included $\lambda \sim 1$ m observations in addition to the canonical $\lambda \sim 20$ cm measurements. Where such pairs of observations exist, papers reporting different populations have been published with particular emphasis on comparing them with longer wavelength observations available in the literature in order to study pulsar emission beam structures: a group of pulsars studied at ~7 cm wavelength (Olszanski et al., 2022); a group of "B" pulsars common to both the Gould & Lyne (1998) and LOFAR High Band (Bilous et al., 2016) surveys (Rankin et al., 2023); and a third group further paralleling the latter survey (Wahl et al., 2023). Here, we provide polarized profiles and emission-beam models for the 59 pulsars for which only $\lambda \sim 20$ cm observations are available. For many pulsars in this group longer wavelength observations are impossible or meaningless because of high interstellar scattering in the inner Galaxy.

The paper is organized as follows: Section 2 discusses our observational methods and the determination of polarized pulsar profiles, § 3 illustrates the techniques used to determine our interstellar RM values and associated parameters; § 4 discusses the polarized pulsar profiles, their morphological classifications, and the emission-beam models implied by them; while § 5 provides final discussions.

## 2 OBSERVATIONS

The observations were carried out using the Arecibo Telescope in Puerto Rico with its Gregorian feed system, 1100-1700-MHz ("L-band" Wide, $\lambda \sim 20$ cm) or 327-MHz ("P-band", $\lambda \sim 1$ m) receivers. Both the L- and P-band antenna systems used crossed linear dipoles and hence linearly polarized inputs to the spectrometers, but a circular hybrid was inserted during certain periods in the 327-MHz configuration, which instead yielded dual circularly polarized inputs. The systems were calibrated by injecting a 25-Hz linearly-polarized noise signal representing Stokes parameter $U$ within the feed.

Almost all of the L-band observations reported here used four Mock spectrometer[1] backends, providing 86-MHz subbands centered at 1170, 1420, 1520 and 1620 MHz. However, some earlier observations prior to MJD 55300 used the Wideband Arecibo Pulsar Processors (WAPPs[2]) with 100-MHz subbands. The spectrometers digitally split each spectrometer's incoming frequency subband into 64 or more adjacent frequency channels, and the time sequences were resampled into ~1000 pulse-phase bins. The lower three subbands were usually free enough of radio frequency interference (RFI) such that they could be combined to give about 300-MHz bandwidth, nominally at 1400 MHz. At P-band, four Mocks with 12.5-MHz subbands were used across the 50-MHz available bandpass for all later observations; however, those prior to MJD 55200 used a single WAPP sampling a 25-MHz subband.

For each subband, the four Stokes parameters were computed from the 4 input signals' auto- and cross-voltage correlations generated by the associated spectrometer. The four polarized channels were then fitted and corrected for Faraday rotation to avoid signal depolarization across the subband (see §3 for details), various instrumental polarization effects, and interstellar dispersion (in order to avoid temporal pulse smearing). In almost all cases, the frequency subbands were channelized sufficiently to provide temporal sequences that were dispersion-smeared by $\lesssim 1$ milliperiod, Faraday-derotated to less than a few degrees, and sampled at milliperiod intervals.

Aside from single-pulse observations (see § 4), the pulsar fractional pulse rotational phase (ranging from 0 to 1 turns) at the exact time of sampling was calculated and each polarization's frequency-channel flux-density output was accumulated into a phase bin labelled by that polarization, frequency, and its associated pulse phase. Our basic analysis unit, a "subband scan," consists then of four two-dimensional (frequency and pulse-phase) arrays, with each array storing the flux density in one of four polarizations across the full bandwidth and pulse period, integrated over several minutes to increase S/N. A "scan" is defined as the set of several simultaneous subband-scans. For each pulsar observed at L-band, Table 2 gives the pulsar's period, the MJD date, total observing duration in pulses, and number of phase bins.

## 3 FARADAY ROTATION MEASURE DETERMINATIONS

### 3.1 Rotation Measure Methods

Faraday rotation along the line of sight produces a PPA rotation as a function of wavelength, as described by eq.(1). Consequently, it produces a linear depolarization across any wide subband if signals are accumulated without compensation. Conversely, if the Faraday rotation is successfully compensated across the band, then the profile will achieve its maximum possible, nominal, linear polarization. We have used this latter criterion to determine the rotation measures of the pulsars in our observations. The procedure for determination of the observed RM and its uncertainty within a single, $i^{th}$ scan, $\mathrm{RM}_{\mathrm{Obs,\,i}}$, is as follows:

- A rough search is carried out over a large range of trial RM values to find the value that maximizes the normalized linear polarization $L/I$ [$=\sqrt{Q^2+U^2}/I$] across all pulsar phase bins within a subband.
- A finer RM search within the subband is then made, centered on the previously found "rough" value. A Gaussian function is fitted to $L/I$ across the central 11 trial RM values, and the RM at which the Gaussian peaks is then defined as our final "best" observed subband RM.
- The RM values and their uncertainties for each of the $\leq 4$ subbands within a scan are compared and deviant determinations culled.
- A further fit as above including all the remaining subbands provides the overall best RM estimate for a given observation, the $\mathrm{i}^{th}$ scan, $\mathrm{RM}_{\mathrm{Obs,\,i}}$.

Computation of appropriate RM error estimates for each scan proved challenging. We initially obtained uncertainties from the Gaussian fit curvature in a manner analogous to correlation detection in pulsar timing. However, these values proved to be unrealistically small.

Instead, we resorted to an empirical technique in which we divide each several-minute scan into time segments (usually 5), compute the RM as above for each, and then take the r.m.s. of the values as the $\mathrm{i}^{th}$scan's RM error estimate, $\Delta\mathrm{RM}_{\mathrm{Obs,\,i}}$.

We were not alone in finding such disparities between the two means of estimating our errors, and some explanations have emerged. Ramachandran et al. (2004) showed that the apparent RM can vary across a pulse profile due to competition between orthogo-

[1] http://www.naic.edu/ao/scientist-user-portal/astronomy/mock-spectrometer
[2] http://www.naic.edu/~wapp

**Table 1. Table A1** Sample: Newly Measured Pulsar RMs and Comparisons with Older Results.

| | | | | | This Paper | | Previous | Previous Reference |
|---|---|---|---|---|---|---|---|---|
| PSR | $\ell$ (°) | $b$ (°) | DM (pc cm$^{-3}$) | DIST (kpc) | N | RM (rad m$^{-2}$) | RM (rad m$^{-2}$) | |
| J0006+1834 | 108.2 | -43.0 | 11.4 | 0.86 | 2 | $-20.54 \pm 0.35$ | $-20.0 \pm 3.0$ | Ng et al. (2020) |
| B0045+33 | 122.3 | -28.7 | 39.9 | 4.50 | 4 | $-82.15 \pm 0.38$ | $-83.3 \pm 1.0$ | Sobey et al. (2019) |
| J0051+0423 | 123.0 | -58.5 | 13.9 | 1.25 | 2 | $-4.90 \pm 0.19$ | | |
| J0122+1416 | 134.0 | -47.9 | 17.7 | 1.58 | 2 | $-10.40 \pm 0.16$ | | |
| J0137+1654 | 138.4 | -44.6 | 26.1 | 3.06 | 1 | $-16.50 \pm 0.36$ | $-16.8 \pm 1.0$ | Sobey et al. (2019) |
| J0152+0948 | 146.8 | -50.2 | 22.9 | 2.48 | 2 | $1.61 \pm 0.59$ | $1.4 \pm 2.0$ | Sobey et al. (2019) |
| J0158+21 | 143.2 | -39.1 | 20.0 | 1.57 | 2 | $-28.84 \pm 0.34$ | $-29.771 \pm 53.0$ | O'Sullivan et al. (2023) |
| J0229+20 | 151.6 | -36.4 | 27.0 | 2.09 | 1 | $-47 \pm 9$ | | |
| J0241+16 | 158.0 | -39.2 | 16.0 | 0.94 | 1 | $-5.60 \pm 0.28$ | | |
| J0244+1433 | 159.9 | -40.1 | 31.0 | 2.80 | 1 | $-6.20 \pm 0.28$ | | |

Notes: All values, except those in the "This Paper" columns, are taken from Version 1.70 of the ATNF Pulsar Catalogue, unless otherwise indicated. A quoted distance of 25.00 kpc implies that a DM distance determination was attempted, but the Yao et al. (2017) galactic model provided too little electron density along the line of sight to do so. See §3.1 for further details.

nally polarized emission modes, which in turn can inflate the measured RM uncertainty. Two recent studies of polarimetry methods further illuminate the origins of this phenomenon. Ilie et al. (2019) analyzed RMs from different parts of a pulsar's profile, and found both that measurement anomalies were more common in multi-component profiles and that aberrant PPAs were encountered deviating from the Faraday law. Then Oswald et al. (2020) studied RMs derived from observations using their Ultra Wideband Receiver, and found that profile evolution across frequency complicated computation of an average polarized profile and the RM its linear polarization represents. Our observations encountered all of these situations, but in a context less suited to identifying the origins of the anomalies.

Using these methods, 547 scans' $RM_{Obs}$ values and error estimates were obtained at L band on 288 pulsars, and 343 values and errors on 172 pulsars were similarly computed at P band.

We compute values of the *interstellar* rotation measure $RM_i$ for each scan *i*, by compensating for the ionospheric contribution, $RM_{iono,i}$, to the observed RM, with

$$RM_i = RM_{obs,\,i} - RM_{iono,\,i}. \tag{3}$$

The estimated $RM_{iono,i}$ along the line of sight at the observation epoch is calculated with the IRI2001 time-variable electron density model (Bilitza, 2003), and the IGRF-11 geomagnetic model (Finlay et al., 2010), with both updated by recent data. The typical uncertainty in the modeled ionospheric RM, $\Delta RM_{iono}$, was determined to be $\sim$0.2 rad m$^{-2}$ via analysis of an intensive set of $\sim$ 400 RM measurements toward PSR B1913+16 over a two-week period—so the ionospheric error is almost always small with respect to the overall RM uncertainties. Nevertheless, each scan's observed RM uncertainty is added in quadrature with its estimated ionospheric RM uncertainty to yield the scan's interstellar RM uncertainty, $\Delta RM_i$.

Multiple scans' interstellar RM values and errors were often available for a given pulsar, so the values were inspected for outliers and corrected when possible. The remaining scans' RMs and error estimates were then averaged with weights proportional to the inverse-square of their deviations from the mean.

### 3.2 Rotation Measure Results

For each of the 314 pulsars whose RMs we measured, Appendix Table A1 (sample in Table 1) displays its interstellar rotation measure *RM* and uncertainty, and *N*, the number of several-minute scans contributing to these measurements; its galactic coordinates $\ell$ and *b*, distance DIST (see below), and *DM*. All of the values displayed in Table A1, aside from our new RMs and ΔRMs, were taken from Version 1.70 of the ATNF Pulsar Catalogue (Manchester et al., 2005),[3] apart from two pulsars' distances that are discussed next.

As the pulsars' distances play a crucial role in the galactic magnetic field analysis of the companion paper (Curtin, Weisberg & Rankin 2023, *in prep*), we carefully define the criteria for determining their distances as follows. For all but two of the pulsars in Table A1, @ARTICLEhmvd18, @ARTICLEhmvd18, we adopted the "best" pulsar distance from the ATNF Pulsar Catalogue, which is denoted as "DIST" in that Catalogue. This distance is defined as the distance derived from the measured DM and the Yao et al. (2017) electron density model, unless a more accurate distance is available from other techniques such as VLBI parallax measurement, kinematic distance via HI 21 cm absorption, or association with another object of known distance. A stated "DIST" of 25.00 kpc indicates that the Yao et al. (2017) model does not provide enough electron density along the line of sight to account for the observed DM.[4] Previous measurements for comparison come from: Brinkman et al. (2018); Camilo et al. (2012); Hamilton & Lyne (1987); Han et al. (2006, 2018); Johnston et al. (2005, 2007, 2020, 2021); Lorimer et al. (2012); Lynch et al. (2013); McLaughlin et al. (2002); Manchester (1974); Noutsos et al. (2008, 2015); Rand & Lyne (1994); Serylak et al. (2021) and Spiewak et al. (2022).

Note that the above analyses implicitly assume that the true

[3] The latest version of the ATNF Pulsar Catalogue is located online at http://www.atnf.csiro.au/research/pulsar/psrcat.

[4] The two exceptions to these rules were PSR J1147+0829 which does not appear in v1.70 of the ATNF catalogue [and whose basic parameters are from K. Stovall (private communication)], for which we applied the Yao et al. (2017) model in an attempt to determine a DM distance; and PSR J2021+3651 (Abdo et al., 2009), where we replaced the ATNF Catalogue's "DIST" value of 1.80 kpc with the Yao et al. (2017) DM distance estimate of 10.51 kpc. This pulsar has a very high DM and RM for its putative "DIST," which was derived from X-ray measurements of the hydrogen column density along the line of sight (Kirichenko et al., 2015). While a localized electron and magnetic enhancement could explain the disproportionately high DM and RM, we believe it is more likely that the X-ray distance determination, which is less well-established than the DM distance technique, has rendered an unrealistically small distance.

**Table 2.** Observation Information for Pulsar Beam Models.

| Pulsar | P(s) | MJD | $N_{pulses}$ | Bins | Pulsar | P(s) | MJD | $N_{pulses}$ | Bins |
|---|---|---|---|---|---|---|---|---|---|
| **Lband** | | | | | | | | | |
| J1802+0128 | 1.109 | 56768 | 536 | 1024 | J1902+0723 | 0.488 | 52739 | 1026 | 512 |
| J1806+1023 | 0.484 | 56605 | 1238 | 1024 | J1904+0004 | 0.140 | 55982 | 2586 | 1019 |
| J1809–0119 | 0.745 | 56768 | 762 | 1455 | J1904+0738 | 0.209 | 57130 | 6359 | 745 |
| J1832+0029 | 0.534 | 58774 | 3365 | 1068 | J1905+0600 | 0.441 | 57004 | 1127 | 1019 |
| J1843–0000 | 0.880 | 56406 | 1044 | 1146 | J1905+0616 | 0.990 | 57113 | 1025 | 1031 |
| J1844+0036 | 0.461 | 56415 | 1297 | 1024 | J1907+0534 | 1.138 | 55633 | 1035 | 1024 |
| J1844–0030 | 0.641 | 57110 | 1866 | 1252 | J1907+0740 | 0.575 | 57114 | 1994 | 1025 |
| J1845+0623 | 1.422 | 57112 | 1021 | 1024 | J1907+0918 | 0.226 | 57130 | 7289 | 1014 |
| J1848–0023 | 0.538 | 57112 | 2041 | 1049 | J1908+0457 | 0.847 | 57114 | 1411 | 1031 |
| J1848+0604 | 2.219 | 57115 | 1031 | 1056 | J1908+0500 | 0.291 | 57114 | 1927 | 1123 |
| J1849+0127 | 0.542 | 57127 | 3130 | 1058 | J1908+0839 | 0.185 | 57125 | 6465 | 715 |
| J1850–0006 | 2.191 | 57115 | 1044 | 1043 | J1910+0728 | 0.325 | 57133 | 3129 | 1048 |
| J1850–0026 | 0.167 | 57112 | 852 | 512 | J1915+0752 | 2.058 | 57131 | 539 | 1029 |
| J1851–0029 | 0.519 | 57120 | 2307 | 1037 | J1916+0844 | 0.440 | 57125 | 1517 | 1050 |
| J1852+0013 | 0.958 | 57120 | 1665 | 1029 | J1920+1040 | 2.216 | 55637 | 525 | 1024 |
| J1852+0305 | 1.326 | 57115 | 1027 | 512 | J1920+1110 | 0.510 | 57133 | 1946 | 1022 |
| J1853–0004 | 0.101 | 57110 | 3497 | 935 | J1926+2016 | 0.299 | 57126 | 3137 | 512 |
| J1853+0505 | 0.905 | 57113 | 1850 | 256 | J1928+1923 | 0.817 | 57113 | 785 | 1033 |
| J1853+0545 | 0.126 | 57120 | 6324 | 1049 | J1931+1952 | 0.501 | 57113 | 4484 | 1046 |
| J1853+0853 | 3.915 | 57639 | 2076 | 1024 | J1935+2025 | 0.080 | 57112 | 4555 | 1023 |
| J1855+0307 | 0.845 | 57120 | 1058 | 1029 | J1941+2525 | 2.306 | 57133 | 796 | 1125 |
| J1855+0422 | 1.678 | 57120 | 860 | 1048 | J1943+0609 | 0.446 | **Johnston (2021)** | | |
| J1855+0527 | 1.393 | 57130 | 1035 | 1032 | J1946+2535 | 0.515 | 57113 | 180 | 1030 |
| J1856+0102 | 0.620 | 57127 | 2343 | 1034 | J2005+3547 | 0.615 | 56768 | 1033 | 512 |
| J1856+0404 | 0.420 | 57004 | 1065 | 512 | J2005+3552 | 0.308 | 56768 | 1947 | 1203 |
| J1857+0526 | 0.350 | 57110 | 1821 | 1001 | J2009+3326 | 1.438 | 56769 | 437 | 512 |
| J1858+0241 | 4.693 | 57568 | 1182 | 1024 | J2011+3331 | 0.932 | 57982 | 888 | 1024 |
| J1900–0051 | 0.385 | 57004 | 1032 | 1071 | J2021+3651 | 0.104 | 56585 | 6316 | 101 |
| J1901+0413 | 2.663 | 57004 | 509 | 1300 | J2030+3641 | 0.200 | 57940 | 1811 | 256 |
| **Pband** | | | | | | | | | |
| J0158+2106 | 0.505 | 57844 | 2374 | 256 | J0229+2258 | 0.807 | 56927 | 1115 | 256 |
| J0241+1604 | 1.545 | 56927 | 1029 | 256 | J2139+00 | 0.312 | 57568 | 2737 | 1024 |

Note: The first three Pband pulsars appear in the current ATNF Pulsar Catalogue as J0158+21, J0229+22 and J0241+16.

interstellar RM is constant over time, whereas it was noted by Weisberg et al. (2004) that the historical record indicates that some pulsars' RMs are time-variable. As we did not encounter any secular trends in our measured RMs, we believe that our sample of pulsars do possess temporally constant RMs, aside from the possibility that single-point outliers might represent a real phenomenon akin to (and conceivably related to) "extreme scattering events" (Fiedler et al., 1987), which are caused by compact regions of greatly enhanced electron density briefly intercepting the line of sight.

## 4 EMISSION-BEAM MODELS FOR AN AO INNER GALAXY POPULATION

We also present polarimetric profiles for the 62 pulsars in Table 2. Most lie toward the inner Galaxy within the Arecibo sky and were observed only at 1.4 GHz. This work continues the efforts begun in Olszanski et al. (2022) and continued in Rankin et al. (2023) and Wahl et al. (2023) on Arecibo pulsars and Rankin (2022) on those elsewhere. Almost all of this group have large dispersion measures and some clear scattering "tails" at 1.4 GHz so observations at lower frequencies were either difficult or useless for any examination of their beam geometry. In a few cases observations at lower frequencies are available elsewhere, generally in the compendia of Bhat et al. (2004), Camilo & Nice (1995), Han et al. (2009), or McEwen et al. (2020), and Johnston & Kerr (2018) provide confirmatory polarimetry for many of the objects. Overall, we have made every effort to locate useful lower frequency profiles, sometimes in the discovery papers. Table 2 gives the MJDs and observational properties of these profiles.

The observations and geometrical models of these 59 pulsars are presented in Table B2 of the Appendix (sample in Table 4). Table B1 (sample in Table 3) gives the physical parameters that can be computed from the period $P$ and spindown $\dot{P}$—that is, the spindown energy $\dot{E}$, spindown age $\tau$, surface magnetic field $B_{surf}$, the acceleration parameter $B_{12}/P^2$ and Beskin *et al.*'s parameter $1/Q$.

### 4.1 Individual Pulsars

Here, our profile measurements, (compiled into Table B2), follow directly from the 1.4-GHz polarized profiles in Figs. B1–B8. Fol-lowing the analysis procedures of Rankin (1993b,a); hereafter ET VI), we have wherever possible measured outside half-power (3 db) widths of the conal components or pairs, while also estimating the sweep rate $R$ of the linear polarization angle (PPA). Where core components can be identified, their 3-db widths facilitate estimation of the magnetic colatitude $\alpha$. We then use the measurement to model the core and conal beam geometry, seeking both a profile classification that describes how our sightline encounters the cone/double-cone beam system and a quantitative emission-beam-geometry model that can include the usual beam dimensions that scale to the angular size of the magnetic polar cap. The model results are given in Table B2.

**Table 3. Table B1** Sample: Pulsar Parameters for Beam Models.

| Pulsar (B1950) | P (s) | $\dot{P}$ (s/s) | $\dot{E}$ (ergs/s) | $\tau$ (yrs) | $B_{surf}$ (G) | $B_{12}$ $/P^2$ | 1/Q |
|---|---|---|---|---|---|---|---|
| J1802+0128 | 1.1085 | 4.22E-15 | 1.23E+32 | 4.16E+06 | 2.18E+12 | 1.8 | 0.8 |
| J1806+1023 | 0.4843 | 5.73E-17 | 2.00E+31 | 1.34E+08 | 1.68E+11 | 0.7 | 0.4 |
| J1809–0119 | 0.7450 | 2.29E-15 | 2.20E+32 | 5.15E+06 | 1.32E+12 | 2.4 | 1.0 |
| J1832+0029 | 0.5339 | 1.51E-15 | 3.90E+32 | 5.60E+06 | 9.09E+11 | 3.2 | 1.2 |
| J1843–0000 | 0.8803 | 7.79E-15 | 4.50E+32 | 1.79E+06 | 2.65E+12 | 3.4 | 1.3 |
| J1844+0036 | 0.4605 | 3.20E-15 | 1.29E+33 | 4.56E+06 | 1.23E+12 | 5.8 | 1.9 |
| J1844–0030 | 0.6411 | 6.08E-15 | 9.10E+32 | 1.67E+06 | 2.00E+12 | 4.9 | 1.7 |
| J1845+0623 | 1.4217 | 5.46E-16 | 7.50E+30 | 4.13E+07 | 8.91E+11 | 0.4 | 0.3 |
| J1848–0023 | 0.5376 | 1.61E-15 | 4.10E+32 | 5.29E+06 | 9.41E+11 | 3.3 | 1.2 |
| J1849+0127 | 0.5422 | 2.80E-14 | 6.90E+33 | 3.07E+05 | 3.94E+12 | 13.4 | 3.7 |

### 4.2 Core and Conal Beams

A full recent discussion of the core/double-cone beam model and its use in computing geometric beam models is given in Rankin (2022).

Canonical pulsar average profiles are observed to have up to five components (Rankin, 1983, ET VI), leading to the conception of the core/double-cone beam model (Backer, 1976). Pulsar profiles then divide into two families depending on whether core or conal emission is dominant at about 1 GHz. Core single $\mathbf{S}_t$ profiles consist of an isolated core component, often flanked by a pair of outriding conal components at high frequency, triple **T** profiles show a core and conal component pair over a wide band, or five-component **M** profiles have a central core component flanked by both an inner and outer pair of conal components.

By contrast, conal profiles can be single $\mathbf{S}_d$ or double **D** when a single cone is involved, or triple c**T** or quadruple c**Q** when the sightline encounters both conal beams. Outer cones tend to have an increasing radius with wavelength, while inner cones tend to show little spectral variation. Periodic modulation often associated with subpulse "drift" is a usual property of conal emission and assists in defining a pulsar's beam configuration (*e.g.,* Rankin, 1986).

Profile classes tend to evolve with frequency in characteristic ways: $\mathbf{S}_t$ profiles often show conal outriders at high frequency, whereas $\mathbf{S}_d$ profiles often broaden and bifurcate at low frequency. **T** profiles tend to show their three components over a broad band, but their relative intensities can change greatly. **M** profiles usually show their five components most clearly at meter wavelengths, while at high frequency they become conflated into a "boxy" form, and at low frequency they become triple because the inner cone often weakens relative to the outer one.

Application of spherical geometry to the measured profile dimensions provides a means of computing the angular beam dimensions—resulting in a quantitative emission-beam model for a given pulsar. Two key angles describing the geometry are the magnetic colatitude (angle between the rotation and magnetic axes) $\alpha$ and the sightline-circle radius (the angle between the rotation axis and the observer's sightline) $\zeta$, where the sightline impact angle $\beta = \zeta - \alpha$.[5] The three beams are found to have particular angular dimensions at 1 GHz in terms of a pulsar's polar cap angular diameter, $\Delta_{PC} = 2.45°P^{-1/2}$ (Rankin, 1990, ET IV). The outside half-power radii of the inner and outer cones, $\rho_i$ and $\rho_o$ are $4.33°P^{-1/2}$ and $5.75°P^{-1/2}$ (ET VIb).

$\alpha$ can be estimated from the core-component width when present, as its half-power width at 1 GHz, $W_{\rm core}$ has been shown to scale as $\Delta_{\rm PC}/\sin\alpha$ (ET IV). The sightline impact angle $\beta$ can then be estimated from the polarization position angle (PPA) sweep rate $R$ [$=|d\chi/d\varphi| = \sin\alpha/\sin\beta$]. Conal beam radii can similarly be estimated from the outside half-power width of a conal component or conal component pair at 1 GHz $W_{\rm cone}$ together with $\alpha$ and $\beta$ using eq.(4) in ET VIa:

$$\rho_{i,o} = \cos^{-1}[\cos\beta - 2\sin\alpha\sin\zeta\sin^2(W_{i,o}/4)] \quad (4)$$

where $W_{i,o}$ is the total half-power width of the inner or outer conal component or pair measured in degrees longitude. The characteristic height of the emission can then be computed assuming dipolarity using eq.(6).

The outflowing plasma responsible for a pulsar's emission is partly or fully generated by a polar "gap" (Ruderman & Sutherland, 1975), just above the stellar surface. Timokhin & Harding (2015) find that this plasma is generated in one of two pair-formation-front (PFF) configurations: for the younger, energetic part of the pulsar population, pairs are created at some 100 m above the polar cap in a central, uniform (1-D) gap potential—thus a 2-D PFF, but for older pulsars the PFF has a lower, annular shape extending up along the polar fluxtube, thus having a 3-D cup shape.

An approximate boundary between the two PFF geometries places the more energetic pulsars to the top left of the $P$-$\dot{P}$ diagram and those less so at the bottom right (*e.g.,* fig. 1 of Olszanski et al. (2022). Its dependence is $\dot{P} = 3.95\times10^{-15}P^{11/4}$. Pulsars with dominant core emission tend to lie to the upper left of the boundary, while the conal population falls to the lower right. In the parlance of ET VI, the division corresponds to an acceleration potential parameter $B_{12}/P^2$ of about 2.5, which in turn represents an energy loss $\dot{E}$ of $10^{32.5}$ ergs/s. This delineation also squares well with Weltevrede & Johnston (2008)'s observation that high energy pulsars have distinct properties and Basu et al. (2016)'s demonstration that conal drifting occurs only for pulsars with $\dot{E}$ less than about $10^{32}$ ergs/s. Table B1 (sample in Table 3 gives the physical parameters that can be computed from the period $P$ and spindown $\dot{P}$, including the $\dot{E}$ and $B_{12}/P^2$.

### 4.3 Computation of Geometric Models

Two observational quantities underlie the computation of conal radii at each frequency and thus the beam model overall: the conal component width(s) and the polarization position angle (PPA) sweep rate $R$ [$=|d\chi/d\varphi|$]; the former gives the angular size of the conal beam(s) while the latter gives the impact angle $\beta$ [$=\sin\alpha/R$] showing how the sightline crosses the beam(s). Following the analysis procedures of ET VI, we have measured outside conal half-power (3 db or FWHM) widths and half-power core widths wherever possible. The measurements are given in Table B2 (sample in Table 4).

[5] $\alpha$ values are defined between 0° and 90°, as $\alpha$ cannot be distinguished from 180° – $\alpha$ in the parlance of Everett & Weisberg (2001).

**Table 4. Table B2** Sample: Profile Geometry Information

| Pulsar | $P$ (s) | DM (pc/cm$^3$) | Class | $W_c$ (°) | $\alpha$ (°) | $R$ (°/°) | $\beta$ (°) | $W_i$ (°) | $\rho_i$ (°) | $\beta/\rho_i$ (°) | $W_o$ (°) | $\rho_o$ (°) | $\beta/\rho_o$ (°) |
|---|---|---|---|---|---|---|---|---|---|---|---|---|---|
| | | | (1-GHz Geometry) | | | | | (1.4-GHz Beam Sizes) | | | | | |
| J1802+0128 | 1.109 | 97.97 | Sd? | — | 53 | -12 | +3.8 | 3.5 | 4.1 | 0.94 | — | — | — |
| J1806+1023 | 0.484 | 52.03 | cQ/M? | — | 80 | ∞ | 0 | 13 | 6.3 | 0.00 | 16.8 | 8.3 | 0.00 |
| J1809–0119 | 0.745 | 140 | St/T? | 2.9 | **78** | ∞ | 0 | 10 | 5.0 | 0.00 | — | — | — |
| J1832+0029 | 0.534 | 28 | St? | 4.1 | **55** | ∞ | 0 | 14 | 5.7 | 0.00 | — | — | — |
| J1843–0000 | 0.880 | 101.5 | T | 4.5 | **35** | -6.4 | -5.2 | — | — | — | 11.2 | 6.0 | -0.86 |
| J1844+0036 | 0.461 | 345.54 | T | 11 | **20** | -3 | 6.5 | — | — | — | 28 | 8.4 | 0.77 |
| J1844–0030 | 0.641 | 605 | St? | 7.0 | **26** | -8 | 3.1 | — | — | — | — | — | — |
| J1845+0623 | 1.422 | 113 | D? | — | 63 | 18 | 2.8 | 5.0 | 3.6 | 0.78 | — | — | — |
| J1848–0023 | 0.538 | 30.6 | D | — | 60 | -11 | 4.5 | — | — | — | 14.4 | 7.8 | 0.58 |
| J1848+0604 | 2.219 | 242.7 | Sd | — | 49 | +24 | -1.8 | 6.1 | 2.9 | -0.62 | — | — | — |

These provide the bases for computing geometrical beaming models for each pulsar, which are also shown in Table B2. $W_c$, $\alpha$, $R$ and $\beta$ are the estimated 1-GHz core width, the magnetic colatitude, the PPA sweep rate and the sightline impact angle; $W_i$, $W_o$ and $\rho_i$, $\rho_o$ are the respective 1.4-GHz inner and outer conal component widths and the respective beam radii.

Core radiation is found empirically to have a bivariate Gaussian (von Mises) beamform such that its 1-GHz (and often invariant) width measures $\alpha$ but provides no sightline impact angle information. If a pulsar has a core component, we attempt to use its width at around 1-GHz to estimate the magnetic colatitude $\alpha$, and when this is possible the $\alpha$ value is bolded in Appendix Table B2 (sample in Table 4). $\beta$ is then estimated from $\alpha$ and $R$ as above. The outside half-power (3 db) widths of conal components or pairs are measured, and the spherical geometry above then used to estimate the outside half-power conal beam radii. Where $\alpha$ can be measured, the value is used, when not an $\alpha$ value is estimated by using the established conal radius or characteristic emission height for an inner or outer cone.

Estimating and propagating errors in the profile widths and PPA sweep rate estimations is awkward. Instead, we assume 10% uncertainties in the conal and core width values as well as in $R$. These then produce typical 10% errors in the model dimensions.

## 4.4 Single-Pulse Modulation

Wherever possible we have also explored a given pulsar's emission on a single-pulse basis, first by computing fluctuation spectra and second by examining the characteristics of its individual pulse sequences. In many cases, however, the former were corrupted by interference and the latter too weak to examine closely, making them less useful.

We have then folded these examinations into short Appendix B paragraphs on each pulsar, explaining the rationale for the type of beam model we have applied or giving reasons for not doing so. These explanations are intended to be read in parallel with Table B2. In some cases interesting phenomena were encountered in the process of the examinations, and for a number of pulsars additional comments are made about the pulse sequences including plots to exhibit particular effects.

## 4.5 Core/double-cone Modeling Results

The 59 pulsars here show beam configurations across all of the core/double-cone model classes. Some half of this group have $\dot{E}$ values $\geq 10^{32.5}$ ergs/s that tend to favor core-single $\mathbf{S}_t$ profiles where the core is present and dominant. The remainder tend to have profiles dominated by conal emission—that is, conal single $\mathbf{S}_d$ or double **D** profiles, conal triple c**T** or weak core-cone triple **T** profiles, quadruple c**Q** or five-component **M** profiles. We were able to construct quantitative beam geometry models for most of the pulsars, though some are better determined than others on the basis of the available information. Lack of reliable PPA rate estimates was a limiting factor in a number of cases, either due to low fractional linear polarization or difficulty interpreting it.

## 4.6 Pulsars With Interesting Characteristics

***J1806+1023*** shows a strong 3-$P$ fluctuation feature in its bright trailing conal component. Fluctuation spectra are shown in Fig B9.
***J1842+0257*** emits a complex mixture of drifting-like emission and long nulls; see Fig. B10.
***J1920+1040***: This pulsar's single pulses come in bursts of 20-30 pulses with deep nulls of usually longer duration in between. Within the bursts the subpulses show a 2.5-$P$ drift modulation. Radar RFI can be seen as well around pulse 339 in Fig. B11.
***J1920+1110***: FLARE or RFI in Fig. B12? The pulsar emits steadily up to about pulse 1470 and then tails off into a long null for the remainder of the 1946-pulse observation. Just before null interval, however, (pulse 1450) we see bright emission around the (unseen) interpulse longitude, and two pulses later (1452) bright emission before and at the main pulse longitude. These two events are the only bright emission in the observation.

# 5 DISCUSSION

We provide rotation-measure values and error estimates in Appendix A for several hundred radio pulsars, many new measurements for weaker pulsars needing Arecibo's great sensitivity and some remeasurements either to confirm or improve previous values. Many values reflect multiple observations at L-band, P-band or both. All have been processed in a manner that accounts for the ionospheric part of the Faraday rotation in a consistent manner, so the values reflect only the RM of the Galactic ISM.

The above RM values together with various other determinations will be used to explore the Galactic magnetic field configuration in a companion paper (Curtin, Weisberg & Rankin 2023, *in prep*).

We also present Arecibo polarized average profiles and analyses in Appendix B aimed at understanding the emission-beam geometries of the about 60 pulsars for which we have Arecibo observations only at L band. This population has small Galactic latitudes $b$, all less than 5° and most around 1°. This reflects that the major part of it in the 18-20 hour right ascension range (some 30° to 75° Galactic longitude) are pulsars at substantial distances, typically 5 to 15 kpc. Thus it is among these pulsars that we encounter both the largest DM and RM values among all of our observations—that is, values up to about 1000 units in each. And it is only among this group that we find pulsars with scattering tails even at L band, rendering them useless for geometric beam modeling.

Regarding the geometric modeling, the 59 pulsars in this population are especially challenging because a) only a few have useful meter-wavelength published profiles to compare, b) nor are there higher frequency profiles available that might show conal outriders, c) about a third are high $\dot{E}$ objects whose profiles are often difficult to interpret, d) most have only profile information so we cannot assess the pulse-train modulation, and e) in a number the component structure has been conflated by scattering. In fact, 1.4-GHz profiles tend to provide the fewest hints about the emission-beam structure, because they tend to lack both clear conal outriders that signal core emission and the better resolved conal structure revealed at lower frequencies as such profiles broaden. Nonetheless, we have made a strong effort that may be taken only as "educated" guesses once further information becomes available.

## 6 OBSERVATIONAL DATA AVAILABILITY

All measured RMs are available in a machine-readable ascii table available through the MNRAS website. The profiles will be available on the European Pulsar Network download site ( http://www.epta.eu.org/epndb/), and the pulse sequences can be obtained by corresponding with the lead author.

## ACKNOWLEDGMENTS

Much of the work was made possible by support from the US NSF grants 18-14397 (JR); and 13-12843 (JMW). The Arecibo Observatory was recently operated by the University of Central Florida. A.P.C. is a Vanier Canada Graduate Scholar.

## APPENDIX A: ARECIBO ROTATION MEASURES FOR 315 PULSARS

Table A1: Newly Measured Pulsar RMs and Comparisons with Older Results.

| | | | | | This Paper | | **Previous** | **Previous Reference** |
|---|---|---|---|---|---|---|---|---|
| PSR | $\ell$ (°) | $b$ (°) | DM (pc cm$^{-3}$) | DIST (kpc) | N | RM (rad m$^{-2}$) | RM (rad m$^{-2}$) | |
| J0006+1834 | 108.2 | -43.0 | 11.4 | 0.86 | 2 | $-20.54\pm0.35$ | $-20.0\pm3.0$ | Ng et al. (2020) |
| B0045+33 | 122.3 | -28.7 | 39.9 | 4.50 | 4 | $-82.15\pm0.38$ | $-83.3\pm1.0$ | Sobey et al. (2019) |
| J0051+0423 | 123.0 | -58.5 | 13.9 | 1.25 | 2 | $-4.90\pm0.19$ | | |
| J0122+1416 | 134.0 | -47.9 | 17.7 | 1.58 | 2 | $-10.40\pm0.16$ | | |
| J0137+1654 | 138.4 | -44.6 | 26.1 | 3.06 | 1 | $-16.50\pm0.36$ | $-16.8\pm1.0$ | Sobey et al. (2019) |
| J0152+0948 | 146.8 | -50.2 | 22.9 | 2.48 | 2 | $1.61\pm0.59$ | $1.4\pm2.0$ | Sobey et al. (2019) |
| J0158+2106[a] | 143.2 | -39.1 | 20.0 | 1.57 | 2 | $-28.84\pm0.34$ | $-29.771\pm53.0$ | O'Sullivan et al. (2023) |
| J0229+2058[a] | 151.6 | -36.4 | 27.0 | 2.09 | 1 | $-47\pm9$ | | |
| J0241+1604[a] | 158.0 | -39.2 | 16.0 | 0.94 | 1 | $-5.60\pm0.28$ | | |
| J0244+1433[a] | 159.9 | -40.1 | 31.0 | 2.80 | 1 | $-6.20\pm0.28$ | | |
| B0301+19 | 161.1 | -33.3 | 15.7 | 0.74 | 6 | $-7.62\pm0.17$ | $-8.43\pm6.0$ | Sobey et al. (2019) |
| J0329+1654 | 168.5 | -31.7 | 40.8 | 2.56 | 4 | $5.45\pm0.65$ | | |
| J0337+1715 | 170.0 | -30.0 | 21.3 | 1.30 | 1 | $29.00\pm0.73$ | | |
| J0348+0432 | 183.3 | -36.8 | 40.5 | 2.10 | 1 | $36\pm20$ | $48.27\pm26.0$ | Spiewak et al. (2022) |
| J0417+35 | 163.2 | -10.5 | 48.5 | 1.48 | 4 | $42.40\pm0.22$ | $42.0\pm1.0$ | Sobey et al. (2019) |
| J0435+2749 | 171.8 | -13.1 | 53.2 | 1.52 | 7 | $2.02\pm0.36$ | $0.96\pm9.0$ | Sobey et al. (2019) |
| J0457+2333[a] | 178.3 | -12.0 | 59.0 | 1.52 | 3 | $-53.29\pm0.84$ | | |
| J0517+2212 | 182.2 | -9.0 | 18.7 | 0.16 | 9 | $-19.58\pm0.57$ | $-20.2\pm16.0$ | Ng et al. (2020) |
| B0523+11 | 192.7 | -13.2 | 79.4 | 1.84 | 3 | $20.79\pm0.66$ | $20.1\pm2.0$ | Sobey et al. (2019) |
| B0525+21 | 183.9 | -6.9 | 50.9 | 1.22 | 6 | $-38.92\pm0.17$ | $-40.02\pm7.0$ | Sobey et al. (2019) |
| J0533+0402 | 200.1 | -15.3 | 83.7 | 2.17 | 2 | $-60.67\pm0.29$ | $-71.0\pm11.0$ | Han et al. (2018) |
| J0538+2817 | 179.7 | -1.7 | 39.6 | 1.30 | 4 | $39.9\pm0.9$ | $39.0\pm3.0$ | Ng et al. (2020) |
| J0540+3207 | 176.7 | 0.8 | 62.0 | 1.42 | 4 | $13.17\pm0.16$ | $13.7\pm18.0$ | Ng et al. (2020) |
| B0540+23 | 184.5 | -3.4 | 77.7 | 1.56 | 5 | $3.25\pm0.43$ | $2.85\pm10.0$ | Sobey et al. (2019) |
| J0546+2441 | 183.7 | -2.0 | 73.8 | 1.52 | 4 | $21.44\pm0.21$ | $24.0\pm3.0$ | Ng et al. (2020) |
| J0609+2130 | 189.2 | 1.0 | 38.7 | 0.57 | 2 | $-7.0\pm1.2$ | | |
| B0609+37 | 175.4 | 9.1 | 27.2 | 0.54 | 6 | $33.1\pm1.8$ | $34.51\pm6.0$ | Sobey et al. (2019) |
| J0613+3731 | 175.3 | 9.2 | 19.0 | 0.19 | 1 | $16.20\pm0.22$ | $16.0\pm2.0$ | Ng et al. (2020) |
| B0611+22 | 188.8 | 2.4 | 96.9 | 3.57 | 5 | $67.27\pm0.48$ | $66.0\pm3.0$ | Johnston et al. (2007) |
| J0621+0336 | 206.2 | -5.1 | 72.6 | 1.70 | 2 | $48.81\pm0.26$ | | |
| J0621+1002 | 200.6 | -2.0 | 36.5 | 0.42 | 1 | $44.0\pm2.0$ | $53.2\pm1.0$ | Sobey et al. (2019) |
| J0623+0340 | 206.5 | -4.5 | 54.0 | 1.38 | 2 | $20.11\pm0.46$ | | |
| J0625+10 | 200.9 | -1.0 | 78.0 | 1.64 | 2 | $50.72\pm0.61$ | | |
| J0627+0649 | 204.2 | -2.1 | 86.6 | 1.79 | 3 | $179.03\pm0.30$ | $179.0\pm3.0$ | Ng et al. (2020) |
| J0627+0706 | 203.9 | -2.0 | 138.2 | 2.29 | 5 | $208.52\pm0.62$ | $212.0\pm0.0$ | Brinkman et al. (2018) |
| J0627+16 | 195.8 | 2.1 | 113.0 | 1.93 | 1 | $15.0\pm5.0$ | | |
| J0628+0909 | 202.2 | -0.9 | 88.3 | 1.77 | 3 | $124.8\pm1.7$ | | |
| B0626+24 | 188.8 | 6.2 | 84.2 | 3.12 | 5 | $72.9\pm3.1$ | $75.7\pm9.0$ | Sobey et al. (2019) |
| J0630-0046 | 211.2 | -5.0 | 97.3 | 2.10 | 2 | $39.5\pm1.1$ | $44.0\pm2.0$ | Johnston et al. (2020) |
| J0631+1036 | 201.2 | 0.5 | 125.4 | 2.10 | 7 | $143.66\pm0.61$ | $137.0\pm4.0$ | Noutsos et al. (2008) |
| J0646+0905 | 204.3 | 3.1 | 149.0 | 2.52 | 2 | $-147.24\pm0.45$ | $-149.0\pm2.0$ | Ng et al. (2020) |
| J0647+0913 | 204.2 | 3.3 | 154.7 | 2.67 | 3 | $-123.99\pm0.85$ | | |
| J0658+0022 | 213.4 | 1.7 | 122.0 | 2.36 | 3 | $85.9\pm2.5$ | | |
| B0656+14 | 201.1 | 8.3 | 13.9 | 0.29 | 6 | $23.88\pm0.13$ | $22.73\pm8.0$ | Sobey et al. (2019) |
| J0711+0931 | 206.7 | 8.8 | 46.2 | 1.17 | 2 | $61.7\pm3.2$ | $62.8\pm11.0$ | Ng et al. (2020) |

| | | | | | This Paper | | Previous | Previous Reference |
|---|---|---|---|---|---|---|---|---|
| PSR | $\ell$ (°) | $b$ (°) | DM (pc cm$^{-3}$) | DIST (kpc) | N | RM (rad m$^{-2}$) | RM (rad m$^{-2}$) | |
| B0751+32 | 188.2 | 26.7 | 40.0 | 1.46 | 6 | $5.06\pm0.21$ | $4.732\pm54.0$ | O'Sullivan et al. (2023) |
| J0806+0811[a] | 213.9 | 20.3 | 46.0 | 1.85 | 1 | $-3.0\pm6.0$ | | |
| J0815+0939 | 213.6 | 22.9 | 52.7 | 2.49 | 3 | $53.24\pm0.28$ | $54.0\pm2.0$ | Ng et al. (2020) |
| B0820+02 | 222.0 | 21.2 | 23.7 | 2.50 | 1 | $17.70\pm0.22$ | $13.0\pm7.0$ | Hamilton & Lyne (1987) |
| B0823+26 | 197.0 | 31.7 | 19.5 | 0.50 | 12 | $6.71\pm0.28$ | $5.38\pm6.0$ | Noutsos et al. (2015) |
| B0834+06 | 219.7 | 26.3 | 12.9 | 0.19 | 3 | $25.49\pm0.74$ | $25.32\pm7.0$ | Noutsos et al. (2015) |
| J0843+0719 | 219.4 | 28.2 | 36.6 | 2.07 | 6 | $43.12\pm0.10$ | $40.0\pm4.0$ | Ng et al. (2020) |
| J0848+1640[a] | 210.1 | 33.3 | 38.0 | 2.28 | 4 | $38.38\pm0.59$ | | |
| B0919+06 | 225.4 | 36.4 | 27.3 | 1.10 | 10 | $33.55\pm0.44$ | $29.2\pm3.0$ | Johnston et al. (2005) |
| J0927+2345 | 205.3 | 44.2 | 17.2 | 1.09 | 7 | $9.60\pm0.39$ | $-8.0\pm0.0$ | Brinkman et al. (2018) |
| J0928+0614[a] | 226.9 | 37.6 | 50.0 | 25.00 | 3 | $18.26\pm0.39$ | | |
| B0940+16 | 216.6 | 45.4 | 20.3 | 1.49 | 1 | $18.0\pm1.0$ | $17.36\pm7.0$ | Sobey et al. (2019) |
| J0943+2253 | 207.9 | 47.5 | 27.3 | 3.56 | 2 | $16.99\pm0.67$ | $16.43\pm8.0$ | Sobey et al. (2019) |
| B0943+10 | 225.4 | 43.1 | 15.3 | 0.89 | 15 | $14.51\pm0.15$ | $14.1\pm8.0$ | Sobey et al. (2019) |
| J0947+2740 | 201.1 | 49.4 | 29.1 | 25.00 | 4 | $25.15\pm0.19$ | $23.25\pm6.0$ | Sobey et al. (2019) |
| B0950+08 | 228.9 | 43.7 | 3.0 | 0.26 | 1 | $1.60\pm0.54$ | $-0.66\pm4.0$ | Johnston et al. (2005) |
| J1022+1001 | 231.8 | 51.1 | 10.3 | 0.72 | 6 | $2.86\pm0.59$ | $1.39\pm5.0$ | Noutsos et al. (2015) |
| J1046+0304 | 246.4 | 51.7 | 25.3 | 5.79 | 3 | $-1.55\pm0.30$ | $-1.1\pm41.0$ | Han et al. (2018) |
| B1133+16 | 241.9 | 69.2 | 4.8 | 0.37 | 3 | $4.76\pm0.31$ | $3.97\pm7.0$ | Noutsos et al. (2015) |
| J1147+0829[b] | 260.7 | 65.9 | 26.9 | 25.00 | 4 | $13.32\pm0.37$ | | |
| J1238+2152 | 272.8 | 84.0 | 18.0 | 2.56 | 3 | $3.14\pm0.24$ | $4.41\pm6.0$ | Sobey et al. (2019) |
| B1237+25 | 252.4 | 86.5 | 9.3 | 0.84 | 7 | $0.28\pm0.22$ | $-0.12\pm6.0$ | Noutsos et al. (2015) |
| J1246+2253 | 288.8 | 85.6 | 17.8 | 2.46 | 3 | $3.8\pm1.4$ | $3.59\pm9.0$ | Sobey et al. (2019) |
| J1313+0931 | 320.4 | 71.7 | 12.0 | 1.08 | 3 | $-2.5\pm1.7$ | $2.3\pm1.0$ | Sobey et al. (2019) |
| J1404+1159 | 355.1 | 67.1 | 18.5 | 2.19 | 8 | $-0.18\pm0.77$ | $4.0\pm0.0$ | Brinkman et al. (2018) |
| J1503+2111 | 29.1 | 59.3 | 3.3 | 0.24 | 7 | $0.03\pm0.35$ | $-5.0\pm10.0$ | Han et al. (2018) |
| B1530+27 | 43.5 | 54.5 | 14.7 | 1.70 | 8 | $5.16\pm0.24$ | $3.37\pm8.0$ | Sobey et al. (2019) |
| B1534+12 | 19.8 | 48.3 | 11.6 | 1.05 | 2 | $5.90\pm0.19$ | $10.6\pm2.0$ | Weisberg et al. (2004) |
| J1538+2345 | 37.3 | 52.4 | 14.9 | 1.31 | 1 | $9.10\pm0.63$ | $11.5\pm11.0$ | Ng et al. (2020) |
| B1541+09 | 17.8 | 45.8 | 35.0 | 5.90 | 4 | $17.38\pm0.46$ | $15.95\pm9.0$ | Sobey et al. (2019) |
| J1549+2113 | 34.6 | 49.1 | 24.1 | 3.03 | 1 | $7.0\pm1.0$ | $32.9\pm87.0$ | Han et al. (2018) |
| B1604-00 | 10.7 | 35.5 | 10.7 | 1.09 | 1 | $7.00\pm0.22$ | $6.5\pm10.0$ | Manchester (1974) |
| J1612+2008 | 35.5 | 43.7 | 19.5 | 1.74 | 3 | $23.40\pm0.27$ | $22.26\pm7.0$ | Sobey et al. (2019) |
| B1612+07 | 20.6 | 38.2 | 21.4 | 1.63 | 3 | $29.03\pm0.25$ | $28.0\pm3.0$ | Ng et al. (2020) |
| J1627+1419 | 30.0 | 38.3 | 32.2 | 4.33 | 3 | $20.13\pm0.38$ | $18.61\pm6.0$ | Sobey et al. (2019) |
| B1633+24 | 43.0 | 39.9 | 24.3 | 2.43 | 5 | $21.8\pm1.5$ | $21.97\pm7.0$ | Sobey et al. (2019) |
| J1645+1012 | 27.7 | 32.5 | 36.2 | 3.91 | 4 | $31.59\pm0.52$ | $30.2\pm1.0$ | Sobey et al. (2019) |
| J1649+2533 | 45.6 | 37.1 | 34.5 | 25.00 | 11 | $28.5\pm0.9$ | $29.52\pm6.0$ | Sobey et al. (2019) |
| J1652+2651 | 47.4 | 37.0 | 40.8 | 25.00 | 7 | $33.93\pm0.90$ | $33.71\pm8.0$ | Sobey et al. (2019) |
| J1713+0747 | 28.8 | 25.2 | 15.9 | 1.15 | 1 | $9.80\pm0.36$ | $10.88\pm10.0$ | Spiewak et al. (2022) |
| J1720+2150 | 44.0 | 29.4 | 40.7 | 5.58 | 6 | $55.6\pm2.0$ | $52.94\pm7.0$ | Sobey et al. (2019) |
| B1726-00 | 23.0 | 18.3 | 41.1 | 1.46 | 5 | $24.82\pm0.85$ | $20.0\pm11.0$ | Han et al. (2018) |
| J1739+0612 | 30.3 | 18.9 | 101.5 | 25.00 | 3 | $30.25\pm0.60$ | | |
| J1740+1000 | 34.0 | 20.3 | 23.9 | 1.23 | 5 | $30.81\pm0.71$ | $23.8\pm28.0$ | McLaughlin et al. (2002) |
| B1737+13 | 37.1 | 21.7 | 48.7 | 4.18 | 5 | $67.35\pm0.36$ | $66.65\pm8.0$ | Sobey et al. (2019) |
| J1741+2758 | 52.3 | 26.7 | 29.1 | 2.52 | 4 | $37.1\pm2.0$ | $54.7\pm7.0$ | Sobey et al. (2019) |
| J1746+2245 | 47.3 | 24.0 | 49.9 | 6.83 | 2 | $77.5\pm2.7$ | $72.3\pm1.0$ | Sobey et al. (2019) |
| J1746+2540 | 50.3 | 25.0 | 51.2 | 25.00 | 3 | $94.69\pm0.27$ | $94.75\pm12.0$ | Sobey et al. (2019) |
| J1752+2359 | 49.1 | 23.1 | 36.2 | 3.03 | 8 | $89.10\pm0.52$ | $87.0\pm1.0$ | Sobey et al. (2019) |
| J1756+1822 | 43.8 | 20.2 | 70.8 | 25.00 | 4 | $73.76\pm0.54$ | $70.0\pm0.0$ | Brinkman et al. (2018) |

| | | | | | | This Paper | Previous | Previous Reference |
|---|---|---|---|---|---|---|---|---|
| PSR | $\ell$ (°) | $b$ (°) | DM (pc cm$^{-3}$) | Dist (kpc) | N | RM (rad m$^{-2}$) | RM (rad m$^{-2}$) | |
| J1758+3030 | 56.3 | 24.0 | 35.1 | 3.28 | 4 | 79.5±1.1 | 79.54±12.0 | Sobey et al. (2019) |
| J1802+0128 | 28.6 | 11.6 | 98.0 | 7.68 | 1 | 38.00±0.28 | | |
| B1802+03 | 30.4 | 11.7 | 80.9 | 5.25 | 3 | 36.9±2.1 | 42.0±2.0 | Ng et al. (2020) |
| J1806+1023 | 37.3 | 14.6 | 52.0 | 2.95 | 1 | 31.60±0.63 | | |
| J1807+0756 | 35.1 | 13.3 | 89.3 | 7.93 | 2 | 144.9±1.8 | | |
| J1809-0119 | 27.0 | 8.6 | 140.0 | 11.09 | 1 | 18.80±0.54 | | |
| J1811+0702 | 34.7 | 12.1 | 57.8 | 2.96 | 2 | 139.60±0.32 | 132.3±27.0 | Han et al. (2018) |
| B1810+02 | 30.7 | 9.7 | 104.1 | 7.20 | 3 | −26.6±1.0 | | |
| J1813+1822 | 45.5 | 16.4 | 60.8 | 4.99 | 1 | 120.0±1.0 | | |
| J1814+1130 | 39.2 | 13.3 | 65.0 | 4.15 | 2 | 68.30±0.48 | | |
| J1819+1305 | 41.2 | 12.8 | 64.8 | 4.05 | 2 | 116.37±0.18 | 114.8±18.0 | Han et al. (2018) |
| J1821+1715 | 45.3 | 14.3 | 60.3 | 4.14 | 1 | 110.70±0.28 | 109.36±13.0 | Sobey et al. (2019) |
| J1822+0705 | 36.0 | 9.7 | 62.2 | 2.87 | 2 | 155.4±1.1 | 158.0±23.0 | Han et al. (2018) |
| B1821+05 | 35.0 | 8.9 | 66.8 | 2.00 | 5 | 142.05±0.27 | 145.0±10.0 | Hamilton & Lyne (1987) |
| B1822+00 | 30.0 | 5.9 | 56.6 | 1.86 | 3 | 21.7±2.0 | 21.0±13.0 | Weisberg et al. (2004) |
| J1828+1359 | 43.0 | 11.2 | 56.0 | 2.75 | 1 | 153.0±4.0 | 121.9±44.0 | Han et al. (2018) |
| J1829+0000 | 30.5 | 4.8 | 114.0 | 4.27 | 1 | −53.0±2.0 | −64.0±24.0 | Han et al. (2018) |
| J1832+0029 | 31.2 | 4.4 | 28.3 | 1.04 | 2 | −19.7±3.2 | | |
| B1831-00 | 30.8 | 3.7 | 88.7 | 3.35 | 3 | 48.2±3.6 | | |
| J1837-0045 | 30.7 | 2.7 | 87.0 | 3.15 | 2 | 136.00±0.66 | 131.6±67.0 | Han et al. (2018) |
| J1837+1221 | 42.4 | 8.7 | 100.6 | 6.60 | 2 | 149.53±0.26 | 173.0±24.0 | Noutsos et al. (2008) |
| J1838+1650 | 46.7 | 10.3 | 33.0 | 1.54 | 2 | 73.08±0.79 | 72.65±12.0 | Sobey et al. (2019) |
| B1839+09 | 40.1 | 6.3 | 49.2 | 1.66 | 3 | 52.20±0.43 | 51.2±1.0 | Sobey et al. (2019) |
| J1842+0257 | 34.6 | 3.3 | 148.1 | 5.04 | 1 | 65.30±0.54 | | |
| J1843-0000 | 32.0 | 1.8 | 101.5 | 3.33 | 3 | −43.7±1.0 | −52.9±38.0 | Han et al. (2018) |
| J1843+2024 | 50.4 | 10.8 | 85.3 | 6.03 | 1 | 151.0±1.0 | | |
| J1844+0036[c] | 32.6 | 1.9 | 345.5 | 6.84 | 2 | 329.9±1.2 | | |
| J1844-0030 | 31.7 | 1.3 | 605.0 | 10.44 | 1 | 214.0±2.0 | 173.0±26.0 | Han et al. (2018) |
| B1842+14 | 45.6 | 8.1 | 41.5 | 1.68 | 3 | 118.76±0.18 | 118.65±12.0 | Sobey et al. (2019) |
| J1845+0623 | 37.9 | 4.3 | 113.0 | 4.92 | 1 | 237.6±0.9 | 226.4±51.0 | Han et al. (2018) |
| J1848-0023 | 32.3 | 0.4 | 30.6 | 1.08 | 1 | 16±9 | | |
| J1848+0604 | 38.1 | 3.3 | 242.7 | 12.60 | 1 | −119.0±2.0 | −119.8±33.0 | Han et al. (2018) |
| J1848+0826 | 40.2 | 4.4 | 90.7 | 3.86 | 4 | 272.0±3.2 | | |
| J1849+0127 | 34.0 | 1.0 | 207.3 | 4.69 | 1 | −138.0±2.0 | −165.0±12.0 | Han et al. (2018) |
| J1849+2423 | 54.7 | 11.2 | 62.3 | 3.98 | 2 | 15.60±0.13 | 15.18±6.0 | Sobey et al. (2019) |
| J1850-0006 | 32.8 | 0.1 | 570.0 | 5.63 | 1 | 672.0±3.0 | | |
| J1850-0026 | 32.4 | 0.1 | 947.0 | 6.71 | 1 | 675.40±0.54 | 664.0±1.0 | Serylak et al. (2021) |
| J1850+0026 | 33.2 | 0.4 | 201.4 | 4.05 | 2 | −17.00±0.28 | −5.4±61.0 | Han et al. (2018) |
| B1848+13 | 45.0 | 6.3 | 60.1 | 2.17 | 3 | 156.1±1.1 | 154.9±2.0 | Sobey et al. (2019) |
| J1851-0029 | 32.5 | -0.3 | 510.0 | 5.45 | 1 | 650.0±2.0 | | |
| J1851-0053 | 32.1 | -0.3 | 24.0 | 0.96 | 1 | −11.20±0.45 | −7.0±16.0 | Han et al. (2018) |
| B1848+04 | 36.7 | 2.0 | 115.5 | 4.09 | 2 | 100.0±1.1 | 86.0±15.0 | Han et al. (2018) |
| B1848+12 | 44.5 | 5.9 | 70.6 | 2.64 | 4 | 150.17±0.21 | 149.8±1.0 | Sobey et al. (2019) |
| J1852+0013 | 33.3 | -0.2 | 545.0 | 5.58 | 1 | 454.0±5.0 | | |
| B1849+00 | 33.5 | 0.0 | 787.0 | 8.00 | 1 | 351±62 | 341.0±11.0 | Han et al. (2018) |
| J1852+0305 | 35.8 | 1.2 | 320.0 | 6.39 | 2 | 245.62±0.45 | 264.0±15.0 | Han et al. (2018) |
| J1853-0004 | 33.1 | -0.5 | 437.5 | 5.34 | 1 | 632.30±0.36 | 627.0±4.0 | Johnston et al. (2021) |
| J1853+0505 | 37.6 | 2.0 | 279.0 | 9.13 | 2 | −26.1±7.9 | | |
| J1853+0545 | 38.4 | 2.1 | 198.7 | 6.55 | 2 | 74.97±0.79 | 79.0±18.0 | Han et al. (2018) |
| J1853+0853 | 41.1 | 3.6 | 214.0 | 12.65 | 1 | 597.0±1.0 | | |

| | | | | | This Paper | | Previous | Previous Reference |
|---|---|---|---|---|---|---|---|---|
| PSR | $\ell$ (°) | $b$ (°) | DM (pc cm$^{-3}$) | DIST (kpc) | N | RM (rad m$^{-2}$) | RM (rad m$^{-2}$) | |
| B1852+10 | 42.9 | 4.2 | 207.2 | 13.84 | 2 | $500.4 \pm 1.6$ | $502.0 \pm 25.0$ | Weisberg et al. (2004) |
| J1855+0307 | 36.2 | 0.5 | 402.5 | 5.94 | 2 | $59.9 \pm 1.3$ | $69.2 \pm 46.0$ | Han et al. (2018) |
| J1855+0422 | 37.3 | 1.1 | 438.0 | 9.99 | 1 | $283 \pm 10$ | | |
| J1855+0527 | 38.2 | 1.6 | 362.0 | 11.71 | 1 | $134.20 \pm 0.45$ | | |
| J1856+0102 | 34.4 | -0.6 | 554.0 | 6.54 | 1 | $549.70 \pm 0.54$ | $556.9 \pm 82.0$ | Han et al. (2018) |
| B1853+01 | 34.6 | -0.5 | 96.7 | 3.30 | 1 | $-125.0 \pm 1.0$ | $-122.0 \pm 3.0$ | Serylak et al. (2021) |
| J1856+0404 | 37.1 | 0.7 | 341.3 | 6.21 | 2 | $-123.0 \pm 1.7$ | $-141.0 \pm 20.0$ | Han et al. (2018) |
| B1854+00 | 34.4 | -0.8 | 82.4 | 2.51 | 2 | $92.53 \pm 0.53$ | $104.0 \pm 19.0$ | Weisberg et al. (2004) |
| B1855+02 | 35.6 | -0.4 | 506.8 | 8.00 | 1 | $420 \pm 9$ | $423.0 \pm 21.0$ | Han et al. (2006) |
| J1857+0526 | 38.4 | 1.2 | 466.4 | 12.26 | 2 | $957.59 \pm 0.47$ | $951.3 \pm 71.0$ | Han et al. (2018) |
| J1858+0241 | 36.2 | -0.4 | 336.0 | 5.15 | 1 | $-34.0 \pm 2.0$ | | |
| J1859+1526 | 47.6 | 5.2 | 97.5 | 4.12 | 4 | $333.71 \pm 0.47$ | $317.0 \pm 10.0$ | Han et al. (2018) |
| J1900-0051 | 33.2 | -2.5 | 136.8 | 4.23 | 1 | $234.0 \pm 2.0$ | $231.4 \pm 88.0$ | Han et al. (2018) |
| J1900+30 | 61.8 | 11.8 | 71.8 | 6.84 | 1 | $121.60 \pm 0.45$ | $121.0 \pm 2.0$ | Ng et al. (2020) |
| B1859+01 | 35.8 | -1.4 | 105.4 | 3.23 | 1 | $-34.70 \pm 0.54$ | $-122.0 \pm 9.0$ | Han et al. (2006) |
| B1859+03 | 37.2 | -0.6 | 402.1 | 7.00 | 3 | $-240.64 \pm 0.78$ | $-237.4 \pm 15.0$ | Hamilton & Lyne (1987) |
| J1901+0413 | 37.8 | -0.2 | 352.0 | 5.34 | 1 | $217.50 \pm 0.54$ | $213.6 \pm 56.0$ | Han et al. (2018) |
| B1859+07 | 40.6 | 1.1 | 252.8 | 3.40 | 5 | $275.44 \pm 0.83$ | $282.0 \pm 13.0$ | Rand & Lyne (1994) |
| J1901+1306 | 45.8 | 3.7 | 75.1 | 2.50 | 2 | $236.23 \pm 0.46$ | | |
| B1900+05 | 39.5 | 0.2 | 177.5 | 3.60 | 1 | $-90.3 \pm 0.9$ | $-113.0 \pm 11.0$ | Hamilton & Lyne (1987) |
| B1900+06 | 39.8 | 0.3 | 502.9 | 7.00 | 2 | $548.0 \pm 1.2$ | $552.6 \pm 59.0$ | Han et al. (2018) |
| J1902+0723 | 40.7 | 1.0 | 105.0 | 3.33 | 1 | $-252 \pm 10$ | $-272.0 \pm 30.0$ | Han et al. (2018) |
| B1900+01 | 35.7 | -2.0 | 245.2 | 3.30 | 3 | $56.4 \pm 8.7$ | $72.3 \pm 10.0$ | Hamilton & Lyne (1987) |
| J1903+2225 | 54.4 | 7.4 | 109.2 | 5.76 | 3 | $52.44 \pm 0.81$ | $57.7 \pm 50.0$ | Han et al. (2018) |
| J1904+0004 | 34.5 | -2.8 | 233.6 | 6.36 | 1 | $300.90 \pm 0.36$ | $289.0 \pm 6.0$ | Noutsos et al. (2008) |
| J1904+0738 | 41.2 | 0.7 | 278.3 | 6.15 | 1 | $141.0 \pm 2.0$ | | |
| B1901+10 | 43.4 | 1.9 | 135.0 | 5.50 | 1 | $-100.0 \pm 2.0$ | $-98.1 \pm 86.0$ | Han et al. (2018) |
| B1902-01 | 33.7 | -3.6 | 229.1 | 7.64 | 2 | $136.4 \pm 1.1$ | $141.8 \pm 60.0$ | Han et al. (2018) |
| J1905+0600 | 39.8 | -0.3 | 730.1 | 8.80 | 1 | $1086.0 \pm 4.0$ | | |
| J1905+0616 | 40.1 | -0.2 | 256.1 | 4.95 | 1 | $150.80 \pm 0.45$ | $136.5 \pm 61.0$ | Han et al. (2018) |
| B1903+07 | 40.9 | 0.1 | 245.3 | 4.98 | 2 | $276.12 \pm 0.20$ | $272.7 \pm 44.0$ | Han et al. (2018) |
| B1904+06 | 40.6 | -0.3 | 472.8 | 7.00 | 2 | $373.73 \pm 0.35$ | $371.5 \pm 23.0$ | Han et al. (2018) |
| J1906+1854 | 51.5 | 5.3 | 156.8 | 6.72 | 3 | $399.69 \pm 0.28$ | $388.0 \pm 10.0$ | Han et al. (2018) |
| J1907+0534 | 39.7 | -1.0 | 524.0 | 11.84 | 2 | $609 \pm 65$ | | |
| J1907+0740 | 41.6 | -0.1 | 332.0 | 5.81 | 1 | $581.80 \pm 0.73$ | $571.0 \pm 15.0$ | Han et al. (2018) |
| J1907+0918 | 43.0 | 0.7 | 357.9 | 8.22 | 1 | $688.0 \pm 1.0$ | $688.8 \pm 2.0$ | Serylak et al. (2021) |
| J1908+0457 | 39.3 | -1.5 | 360.0 | 11.28 | 1 | $956.10 \pm 0.22$ | $929.0 \pm 30.0$ | Han et al. (2018) |
| J1908+0500 | 39.3 | -1.4 | 201.4 | 5.84 | 1 | $134.70 \pm 0.45$ | $120.6 \pm 66.0$ | Han et al. (2018) |
| J1908+0839 | 42.6 | 0.2 | 512.1 | 8.27 | 1 | $801 \pm 22$ | | |
| J1908+2351 | 56.1 | 7.1 | 101.7 | 5.41 | 4 | $56.50 \pm 0.26$ | | |
| B1907+00 | 35.1 | -4.0 | 112.8 | 4.36 | 2 | $7.26 \pm 0.41$ | $-40.0 \pm 15.0$ | Hamilton & Lyne (1987) |
| B1907+02 | 37.6 | -2.7 | 171.7 | 4.50 | 3 | $269.81 \pm 0.53$ | $254.0 \pm 14.0$ | Han et al. (2018) |
| B1907+10 | 44.8 | 1.0 | 150.0 | 4.80 | 2 | $553.84 \pm 0.48$ | $540.0 \pm 20.0$ | Hamilton & Lyne (1987) |
| J1909+1859 | 51.9 | 4.7 | 64.5 | 2.45 | 4 | $217.7 \pm 1.2$ | $216.7 \pm 28.0$ | Han et al. (2018) |
| B1907+03 | 38.6 | -2.3 | 82.9 | 2.86 | 1 | $-125.0 \pm 1.0$ | $-127.0 \pm 7.0$ | Hamilton & Lyne (1987) |
| J1910+0714 | 41.5 | -0.9 | 124.1 | 3.68 | 4 | $148.28 \pm 0.72$ | $158.0 \pm 12.0$ | Han et al. (2018) |
| J1910+0728 | 41.7 | -0.8 | 283.7 | 6.23 | 1 | $567.60 \pm 0.45$ | $550.0 \pm 11.0$ | Han et al. (2018) |
| B1907+12 | 46.2 | 1.6 | 258.6 | 8.14 | 2 | $997.4 \pm 6.4$ | $978.0 \pm 15.0$ | Weisberg et al. (2004) |
| J1911+1758 | 51.2 | 3.7 | 49.0 | 1.96 | 3 | $142.2 \pm 1.5$ | $148.4 \pm 50.0$ | Han et al. (2018) |
| B1910+10 | 44.8 | 0.2 | 147.0 | 4.15 | 1 | $153 \pm 27$ | $147.3 \pm 69.0$ | Han et al. (2018) |

| | | | | | This Paper | | Previous | Previous Reference |
|---|---|---|---|---|---|---|---|---|
| PSR | $\ell$ (°) | $b$ (°) | DM (pc cm$^{-3}$) | DIST (kpc) | N | RM (rad m$^{-2}$) | RM (rad m$^{-2}$) | |
| B1910+20 | 54.1 | 5.0 | 88.6 | 3.37 | 5 | $148.25 \pm 0.81$ | $148.0 \pm 10.0$ | Hamilton & Lyne (1987) |
| J1912+2525 | 57.9 | 7.0 | 37.8 | 2.15 | 3 | $33.1 \pm 1.0$ | $32.65 \pm 10.0$ | Sobey et al. (2019) |
| J1913+0446 | 39.7 | -2.8 | 109.1 | 4.34 | 2 | $-99.45 \pm 0.34$ | $-97.3 \pm 43.0$ | Han et al. (2018) |
| B1911+09 | 44.0 | -0.6 | 157.0 | 4.34 | 1 | $143.6 \pm 0.9$ | | |
| B1911+13 | 47.9 | 1.6 | 145.1 | 5.26 | 1 | $598.0 \pm 1.0$ | $435.0 \pm 30.0$ | Hamilton & Lyne (1987) |
| J1913+3732 | 69.1 | 12.1 | 72.3 | 7.58 | 1 | $1.40 \pm 0.22$ | | |
| J1914+0219 | 37.6 | -4.0 | 233.8 | 14.12 | 1 | $275.10 \pm 0.22$ | $279.2 \pm 25.0$ | Han et al. (2018) |
| B1911+11 | 45.6 | 0.2 | 100.0 | 3.14 | 1 | $369.0 \pm 2.0$ | $360.9 \pm 72.0$ | Han et al. (2018) |
| J1915+0227 | 37.8 | -4.1 | 192.6 | 10.28 | 1 | $162.0 \pm 1.0$ | | |
| J1915+0738 | 42.5 | -1.8 | 39.0 | 1.40 | 1 | $-2.80 \pm 0.36$ | $-6.8 \pm 15.0$ | Han et al. (2018) |
| J1915+0752 | 42.6 | -1.6 | 105.3 | 3.57 | 1 | $186 \pm 19$ | $211.3 \pm 26.0$ | Han et al. (2018) |
| B1913+10 | 44.7 | -0.7 | 241.7 | 7.00 | 1 | $427.90 \pm 0.22$ | $430.0 \pm 6.0$ | Johnston et al. (2007) |
| B1913+16 | 50.0 | 2.1 | 168.8 | 4.17 | 400 | $357.29 \pm 0.07$ | $364.5 \pm 50.0$ | Han et al. (2018) |
| B1913+167 | 50.6 | 2.5 | 62.6 | 2.20 | 3 | $181.8 \pm 0.9$ | $172.0 \pm 3.0$ | Weisberg et al. (2004) |
| J1916+0844 | 43.5 | -1.5 | 339.4 | 11.01 | 1 | $598.20 \pm 0.63$ | $582.0 \pm 12.0$ | Han et al. (2018) |
| B1914+09 | 44.6 | -1.0 | 61.0 | 1.90 | 1 | $98.10 \pm 0.63$ | $97.0 \pm 1.0$ | Johnston et al. (2007) |
| B1914+13 | 47.6 | 0.5 | 237.0 | 4.50 | 2 | $276.7 \pm 1.4$ | $280.0 \pm 15.0$ | Hamilton & Lyne (1987) |
| B1915+13 | 48.3 | 0.6 | 94.5 | 5.88 | 4 | $228.62 \pm 0.32$ | $233.0 \pm 8.0$ | Hamilton & Lyne (1987) |
| B1915+22 | 55.8 | 4.6 | 134.7 | 4.96 | 1 | $191.40 \pm 0.28$ | $168.0 \pm 24.0$ | Han et al. (2018) |
| B1916+14 | 49.1 | 0.9 | 27.2 | 1.30 | 3 | $-34.65 \pm 0.32$ | $-41.7 \pm 35.0$ | Han et al. (2018) |
| B1917+00 | 36.5 | -6.2 | 90.3 | 5.88 | 5 | $118.65 \pm 0.22$ | $120.0 \pm 7.0$ | Hamilton & Lyne (1987) |
| J1919+0134 | 37.6 | -5.6 | 191.9 | 14.37 | 2 | $53.4 \pm 2.0$ | $47.0 \pm 4.0$ | Noutsos et al. (2008) |
| J1920+1040 | 45.8 | -1.6 | 304.0 | 10.15 | 1 | $772.0 \pm 1.0$ | $733.0 \pm 16.0$ | Han et al. (2018) |
| J1920+1110 | 46.2 | -1.2 | 182.0 | 6.13 | 1 | $646.0 \pm 5.0$ | $650.0 \pm 25.0$ | Han et al. (2018) |
| B1918+26 | 60.1 | 6.0 | 27.7 | 1.72 | 2 | $23.33 \pm 0.62$ | $26.04 \pm 7.0$ | Sobey et al. (2019) |
| B1919+14 | 49.1 | 0.0 | 91.6 | 2.81 | 1 | $160.10 \pm 0.22$ | $164.8 \pm 31.0$ | Han et al. (2018) |
| B1918+19 | 53.9 | 2.7 | 153.8 | 4.32 | 3 | $152.7 \pm 2.3$ | $160.0 \pm 20.0$ | Hamilton & Lyne (1987) |
| B1919+20 | 54.2 | 2.6 | 101.0 | 3.29 | 1 | $140.0 \pm 1.0$ | $128.2 \pm 43.0$ | Han et al. (2018) |
| B1919+21 | 55.8 | 3.5 | 12.4 | 0.30 | 4 | $-16.572 \pm 0.036$ | $-16.99 \pm 5.0$ | Noutsos et al. (2015) |
| B1920+21 | 55.3 | 2.9 | 217.1 | 4.00 | 2 | $276.2 \pm 3.8$ | $282.0 \pm 14.0$ | Hamilton & Lyne (1987) |
| B1921+17 | 51.7 | 1.0 | 142.5 | 3.98 | 1 | $596.0 \pm 3.0$ | $380.0 \pm 220.0$ | Weisberg et al. (2004) |
| B1924+14 | 49.9 | -1.0 | 211.4 | 5.65 | 2 | $252.0 \pm 1.2$ | $249.3 \pm 27.0$ | Han et al. (2018) |
| B1924+16 | 51.9 | 0.1 | 176.9 | 6.00 | 4 | $317.23 \pm 0.52$ | $320.0 \pm 14.0$ | Hamilton & Lyne (1987) |
| J1926+2016 | 54.9 | 1.8 | 247.0 | 5.95 | 1 | $184.0 \pm 1.0$ | | |
| B1925+18 | 53.7 | 1.0 | 254.0 | 5.30 | 1 | $254.20 \pm 0.63$ | $417.0 \pm 70.0$ | Weisberg et al. (2004) |
| B1925+188 | 53.8 | 0.9 | 99.0 | 3.10 | 1 | $103 \pm 12$ | $74.4 \pm 58.0$ | Han et al. (2018) |
| B1925+22 | 57.0 | 2.7 | 180.0 | 6.71 | 2 | $232.23 \pm 0.79$ | $215.7 \pm 75.0$ | Han et al. (2018) |
| J1928+1923 | 54.3 | 1.0 | 476.0 | 10.57 | 1 | $335.5 \pm 1.4$ | | |
| J1931+1952 | 55.1 | 0.5 | 441.0 | 9.62 | 1 | $159.0 \pm 2.0$ | | |
| B1929+10 | 47.4 | -3.9 | 3.2 | 0.31 | 3 | $-6.30 \pm 0.40$ | $-6.87 \pm 2.0$ | Johnston et al. (2005) |
| B1930+22 | 57.4 | 1.6 | 219.2 | 10.90 | 1 | $140.10 \pm 0.22$ | $138.9 \pm 1.0$ | Serylak et al. (2021) |
| B1930+13 | 49.4 | -3.1 | 177.0 | 6.08 | 2 | $-120.00 \pm 0.11$ | $-124.8 \pm 38.0$ | Han et al. (2018) |
| B1931+24 | 59.5 | 2.4 | 106.0 | 4.64 | 3 | $117.0 \pm 1.3$ | $114.8 \pm 20.0$ | Han et al. (2018) |
| J1935+1159 | 48.6 | -4.1 | 188.8 | 8.42 | 2 | $-84.0 \pm 1.3$ | $-83.0 \pm 0.0$ | Brinkman et al. (2018) |
| B1933+16 | 52.4 | -2.1 | 158.5 | 3.70 | 5 | $-1.4 \pm 1.2$ | $-10.2 \pm 3.0$ | Johnston et al. (2005) |
| J1935+2025 | 56.1 | -0.1 | 182.0 | 4.60 | 1 | $19.70 \pm 0.28$ | | |
| B1935+25 | 60.8 | 2.3 | 53.2 | 3.12 | 5 | $27.36 \pm 0.89$ | $26.0 \pm 3.0$ | Johnston et al. (2007) |
| J1938+0650 | 44.4 | -7.1 | 70.8 | 2.83 | 2 | $91.3 \pm 1.6$ | $97.5 \pm 122.0$ | Han et al. (2018) |
| J1941+1026 | 48.0 | -6.2 | 138.9 | 7.97 | 2 | $-117.4 \pm 1.3$ | $-126.3 \pm 64.0$ | Han et al. (2018) |
| J1941+2525 | 61.0 | 1.3 | 314.4 | 11.17 | 2 | $-103.78 \pm 0.32$ | | |

| | | | | | | This Paper | Previous | Previous Reference |
|---|---|---|---|---|---|---|---|---|
| PSR | $\ell$ (°) | $b$ (°) | DM (pc cm$^{-3}$) | DIST (kpc) | N | RM (rad m$^{-2}$) | RM (rad m$^{-2}$) | |
| B1942-00 | 38.6 | -12.3 | 59.7 | 3.16 | 3 | $-54.1\pm1.6$ | $-45.0\pm7.0$ | Hamilton & Lyne (1987) |
| B1944+17 | 55.3 | -3.5 | 16.1 | 0.30 | 3 | $-42.9\pm1.7$ | $-45.34\pm11.0$ | Sobey et al. (2019) |
| B1944+22 | 59.3 | -1.1 | 140.0 | 5.95 | 3 | $6.9\pm1.4$ | $2.0\pm20.0$ | Weisberg et al. (2004) |
| J1946+2535 | 61.8 | 0.3 | 248.8 | 8.30 | 1 | $57.0\pm1.0$ | $57.0\pm3.0$ | Ng et al. (2020) |
| B1946+35 | 70.7 | 5.0 | 129.4 | 7.65 | 4 | $119.69\pm0.34$ | $118.96\pm11.0$ | Sobey et al. (2019) |
| J1951+1123 | 50.0 | -7.7 | 31.3 | 1.51 | 3 | $-50.11\pm0.33$ | $-52.8\pm37.0$ | Han et al. (2018) |
| B1949+14 | 52.5 | -6.6 | 31.5 | 1.59 | 3 | $-20.77\pm0.16$ | $-11.5\pm45.0$ | Han et al. (2018) |
| B1951+32 | 68.8 | 2.8 | 45.0 | 3.00 | 1 | $-184.0\pm1.0$ | $-182.0\pm8.0$ | Weisberg et al. (2004) |
| J1953+1149 | 50.7 | -8.1 | 140.0 | 10.52 | 2 | $-13.5\pm6.0$ | | |
| B1952+29 | 65.9 | 0.8 | 7.9 | 0.54 | 4 | $-15.49\pm0.33$ | $-16.0\pm2.0$ | Ng et al. (2020) |
| B1953+29 | 65.8 | 0.4 | 104.5 | 6.30 | 3 | $4.52\pm0.45$ | $15.3\pm5.0$ | Spiewak et al. (2022) |
| J1956+0838 | 48.3 | -10.3 | 67.1 | 3.54 | 1 | $-107.70\pm0.22$ | $-113.5\pm2.0$ | Sobey et al. (2019) |
| J1957+2831 | 65.5 | -0.2 | 139.0 | 6.89 | 1 | $-41.0\pm1.0$ | $-43.7\pm18.0$ | Han et al. (2018) |
| J1959+3620 | 72.4 | 3.4 | 273.0 | 10.89 | 1 | $61.4\pm1.0$ | | |
| J2002+1637 | 56.0 | -7.5 | 94.6 | 4.90 | 3 | $-40.7\pm1.1$ | $-41.3\pm1.0$ | Sobey et al. (2019) |
| B2000+32 | 69.3 | 0.9 | 142.2 | 6.46 | 4 | $-89.93\pm0.73$ | $-90.2\pm13.0$ | Han et al. (2018) |
| B2002+31 | 69.0 | 0.0 | 234.8 | 8.00 | 5 | $30.4\pm2.1$ | $31.5\pm8.0$ | Ng et al. (2020) |
| J2005+3547 | 72.6 | 2.2 | 401.6 | 13.56 | 1 | $19.0\pm2.0$ | | |
| J2005+3552 | 72.7 | 2.1 | 455.0 | 14.65 | 1 | $211.20\pm0.36$ | | |
| J2007+0809 | 49.2 | -12.8 | 53.4 | 3.04 | 3 | $-132.51\pm0.31$ | $-133.3\pm2.0$ | Sobey et al. (2019) |
| J2007+0910 | 50.2 | -12.4 | 48.7 | 2.63 | 1 | $-79.30\pm0.82$ | $-77.8\pm1.0$ | Sobey et al. (2019) |
| J2008+2513 | 64.1 | -4.1 | 60.6 | 4.03 | 2 | $-122.5\pm1.4$ | $-116.6\pm14.0$ | Han et al. (2018) |
| J2009+3326 | 71.1 | 0.1 | 263.6 | 7.30 | 1 | $12.0\pm2.0$ | | |
| J2010+2845 | 67.2 | -2.5 | 112.5 | 6.58 | 2 | $-233.33\pm0.57$ | $-233.0\pm3.0$ | Ng et al. (2020) |
| J2011+3331 | 71.3 | -0.0 | 298.6 | 8.60 | 1 | $225.0\pm2.0$ | $235.6\pm5.0$ | Ng et al. (2020) |
| J2015+2524 | 65.0 | -5.3 | 13.0 | 0.88 | 4 | $-31.14\pm0.60$ | | |
| J2016+1948 | 60.5 | -8.7 | 33.8 | 2.16 | 2 | $-122.2\pm1.3$ | $-121.0\pm4.0$ | Ng et al. (2020) |
| J2017+2043 | 61.4 | -8.3 | 60.5 | 4.21 | 4 | $-163.6\pm0.9$ | $-163.6\pm1.0$ | Sobey et al. (2019) |
| B2016+28 | 68.1 | -4.0 | 14.2 | 0.98 | 3 | $-28.9\pm2.0$ | $-34.9\pm1.0$ | Sobey et al. (2019) |
| J2018+3431 | 73.0 | -0.8 | 222.3 | 6.64 | 1 | $-9.0\pm5.0$ | | |
| J2021+3651[d] | 75.2 | 0.1 | 367.5 | 10.51 | 2 | $517.3\pm1.8$ | $524.0\pm4.0$ | Abdo et al. (2009) |
| B2020+28 | 68.9 | -4.7 | 24.6 | 2.10 | 6 | $-74.91\pm0.43$ | $-75.04\pm6.0$ | Sobey et al. (2019) |
| B2025+21 | 63.5 | -9.6 | 97.1 | 10.29 | 2 | $-212.5\pm5.7$ | $-210.0\pm2.0$ | Ng et al. (2020) |
| B2027+37 | 76.9 | -0.7 | 190.7 | 5.77 | 4 | $-6.0\pm1.1$ | $-6.0\pm18.0$ | Han et al. (2018) |
| B2028+22 | 64.6 | -9.8 | 71.9 | 6.39 | 2 | $-192.61\pm0.40$ | $-192.0\pm21.0$ | Weisberg et al. (2004) |
| J2030+3641 | 76.1 | -1.4 | 246.0 | 6.95 | 1 | $514.0\pm5.0$ | $514.0\pm1.0$ | Camilo et al. (2012) |
| J2033+0042 | 45.9 | -22.2 | 37.8 | 2.93 | 1 | $-71.40\pm0.22$ | $-71.2\pm22.0$ | Lynch et al. (2013) |
| J2036+2835 | 70.4 | -7.4 | 84.2 | 6.74 | 2 | $-157.27\pm0.31$ | $-158.14\pm12.0$ | Sobey et al. (2019) |
| B2034+19 | 63.2 | -12.7 | 36.9 | 2.79 | 7 | $-109.88\pm0.45$ | $-110.7\pm1.0$ | Sobey et al. (2019) |
| B2035+36 | 76.7 | -2.8 | 93.6 | 4.85 | 3 | $251.93\pm0.54$ | $252.2\pm11.0$ | Han et al. (2018) |
| J2040+1657 | 61.3 | -14.9 | 50.7 | 4.45 | 3 | $-98.60\pm0.25$ | $-98.0\pm2.0$ | Ng et al. (2020) |
| J2043+2740 | 70.6 | -9.2 | 21.0 | 1.48 | 2 | $-96.20\pm0.14$ | $-96.1\pm1.0$ | Serylak et al. (2021) |
| J2045+0912 | 55.4 | -20.3 | 31.8 | 2.46 | 1 | $-86.50\pm0.22$ | $-84.7\pm19.0$ | Ng et al. (2020) |
| B2044+15 | 61.1 | -16.8 | 39.8 | 3.23 | 6 | $-90.18\pm0.19$ | $-90.2\pm1.0$ | Sobey et al. (2019) |
| J2048+2255 | 67.5 | -12.9 | 70.7 | 7.65 | 3 | $-166.8\pm1.7$ | | |
| J2050+1259 | 59.4 | -19.2 | 52.4 | 5.83 | 3 | $-78.47\pm0.34$ | $-80.0\pm0.0$ | Brinkman et al. (2018) |
| J2053+1718 | 63.6 | -17.3 | 27.0 | 2.09 | 3 | $-55.01\pm0.24$ | $-5.0\pm0.0$ | Brinkman et al. (2018) |
| B2053+21 | 67.8 | -14.7 | 36.3 | 3.00 | 5 | $-91.1\pm1.5$ | $-89.4\pm8.0$ | Sobey et al. (2019) |
| B2053+36 | 79.1 | -5.6 | 97.4 | 5.00 | 4 | $-61.1\pm1.2$ | $-63.4\pm3.0$ | Ng et al. (2020) |
| J2111+2106 | 69.4 | -18.2 | 59.3 | 8.41 | 7 | $-75.25\pm0.27$ | $-75.0\pm1.0$ | Sobey et al. (2019) |

| | | | | | | This Paper | Previous | Previous Reference |
|---|---|---|---|---|---|---|---|---|
| PSR | $\ell$ (°) | $b$ (°) | DM (pc cm$^{-3}$) | DIST (kpc) | N | RM (rad m$^{-2}$) | RM (rad m$^{-2}$) | |
| B2110+27 | 75.0 | -14.0 | 25.1 | 1.43 | 5 | $-56.0 \pm 3.6$ | $-60.2 \pm 1.0$ | Sobey et al. (2019) |
| B2113+14 | 64.5 | -23.4 | 56.2 | 25.00 | 3 | $-39.8 \pm 1.7$ | $-39.0 \pm 1.0$ | Sobey et al. (2019) |
| B2122+13 | 65.8 | -25.1 | 30.2 | 2.81 | 4 | $-44.79 \pm 0.44$ | $-43.0 \pm 3.0$ | Ng et al. (2020) |
| J2139+00 | 55.9 | -36.3 | 36.0 | 25.00 | 1 | $8.70 \pm 0.28$ | $11.0 \pm 2.0$ | Ng et al. (2020) |
| J2139+2242 | 75.3 | -21.9 | 44.2 | 4.93 | 4 | $-86.90 \pm 0.24$ | $-87.4 \pm 2.0$ | Sobey et al. (2019) |
| J2151+2315 | 77.8 | -23.5 | 23.6 | 1.91 | 2 | $-18.1 \pm 1.2$ | $112.0 \pm 30.0$ | Han et al. (2018) |
| J2155+2813 | 82.1 | -20.4 | 77.1 | 25.00 | 5 | $-130.73 \pm 0.22$ | $-132.7 \pm 1.0$ | Sobey et al. (2019) |
| J2156+2618 | 81.0 | -22.0 | 48.4 | 6.10 | 3 | $-60.4 \pm 1.2$ | $-72.5 \pm 72.0$ | Han et al. (2018) |
| J2205+1444 | 73.9 | -31.9 | 36.7 | 6.04 | 4 | $-23.8 \pm 1.4$ | $-25.5 \pm 1.0$ | Sobey et al. (2019) |
| B2210+29 | 86.1 | -21.7 | 74.5 | 25.00 | 4 | $-167.94 \pm 0.43$ | $-168.7 \pm 1.0$ | Sobey et al. (2019) |
| J2215+1538 | 76.8 | -32.9 | 29.2 | 3.14 | 5 | $-20.4 \pm 1.1$ | $-20.02 \pm 9.0$ | Sobey et al. (2019) |
| J2222+2923 | 88.0 | -23.2 | 49.4 | 7.63 | 3 | $-96.78 \pm 0.34$ | $-95.7 \pm 1.0$ | Sobey et al. (2019) |
| J2234+2114 | 85.1 | -31.4 | 35.1 | 4.47 | 4 | $-94.7 \pm 1.3$ | $-93.7 \pm 18.0$ | Ng et al. (2020) |
| J2243+1518 | 82.8 | -37.4 | 42.1 | 25.00 | 3 | $-35.49 \pm 0.44$ | $-35.5 \pm 5.0$ | Ng et al. (2020) |
| J2248-0101 | 69.3 | -50.6 | 29.1 | 4.00 | 2 | $12 \pm 13$ | $33.0 \pm 12.0$ | Han et al. (2006) |
| J2253+1516 | 85.3 | -38.8 | 29.2 | 3.81 | 4 | $-31.16 \pm 0.39$ | $-31.0 \pm 1.0$ | Sobey et al. (2019) |
| B2303+30 | 97.7 | -26.7 | 49.6 | 4.35 | 4 | $-85.30 \pm 0.78$ | $-87.0 \pm 1.0$ | Sobey et al. (2019) |
| J2307+2225 | 93.6 | -34.5 | 7.0 | 0.49 | 4 | $-8.1 \pm 2.0$ | $-11.6 \pm 11.0$ | Han et al. (2018) |
| B2315+21 | 95.8 | -36.1 | 20.9 | 1.96 | 4 | $-36.2 \pm 2.3$ | $-37.698 \pm 51.0$ | O'Sullivan et al. (2023) |
| J2355+2246 | 106.5 | -38.3 | 23.1 | 2.16 | 3 | $-53.98 \pm 0.65$ | | |

Notes: All values, except those in the "This Paper" columns, are taken from Version 1.70 of the ATNF Pulsar Catalogue, unless otherwise indicated. A quoted distance of 25.00 kpc implies that a DM distance determination was attempted, but the Yao et al. (2017) galactic model provided too little electron density along the line of sight to do so. See §3.1 for further details.

[a] Position and timing information provided by Kevin Stovall (private communication); these appear in the ATNF Catalogue (1.70) with names lacking the final two digits. [b] Position and timing information provided by Kevin Stovall as above. This pulsar does not appear in the ATNF Catalogue (1.70). [c] This pulsar appears in the ANTF Catalogue (1.70) as J1844+00; we use its (inaccurate) declination to rename it J1844+0036, thereby distinguishing its designation from those of other nearby pulsars [d] Distance from Yao et al.(2017). See §3.1 for further details.

## APPENDIX B: PULSAR BEAM GEOMETRY TABLES, MODELS AND NOTES

Table B1: Pulsar Parameters

| Pulsar (B1950) | P (s) | $\dot{P}$ (s/s) | $\dot{E}$ (ergs/s) | $\tau$ (yrs) | $B_{surf}$ (G) | $B_{12}/P^2$ | 1/Q |
|---|---|---|---|---|---|---|---|
| J1802+0128 | 1.1085 | 4.22E-15 | 1.23E+32 | 4.16E+06 | 2.18E+12 | 1.8 | 0.8 |
| J1806+1023 | 0.4843 | 5.73E-17 | 2.00E+31 | 1.34E+08 | 1.68E+11 | 0.7 | 0.4 |
| J1809–0119 | 0.7450 | 2.29E-15 | 2.20E+32 | 5.15E+06 | 1.32E+12 | 2.4 | 1.0 |
| J1832+0029 | 0.5339 | 1.51E-15 | 3.90E+32 | 5.60E+06 | 9.09E+11 | 3.2 | 1.2 |
| J1843–0000 | 0.8803 | 7.79E-15 | 4.50E+32 | 1.79E+06 | 2.65E+12 | 3.4 | 1.3 |
| J1844+0036 | 0.4605 | 3.20E-15 | 1.29E+33 | 4.56E+06 | 1.23E+12 | 5.8 | 1.9 |
| J1844–0030 | 0.6411 | 6.08E-15 | 9.10E+32 | 1.67E+06 | 2.00E+12 | 4.9 | 1.7 |
| J1845+0623 | 1.4217 | 5.46E-16 | 7.50E+30 | 4.13E+07 | 8.91E+11 | 0.4 | 0.3 |
| J1848–0023 | 0.5376 | 1.61E-15 | 4.10E+32 | 5.29E+06 | 9.41E+11 | 3.3 | 1.2 |
| J1848+0604 | 2.2186 | 3.74E-15 | 1.40E+31 | 9.41E+06 | 2.91E+12 | 0.6 | 0.4 |
| J1849+0127 | 0.5422 | 2.80E-14 | 6.90E+33 | 3.07E+05 | 3.94E+12 | 13.4 | 3.7 |
| J1850–0006 | 2.1915 | 4.32E-15 | 1.60E+31 | 8.04E+06 | 3.11E+12 | 0.6 | 0.4 |
| J1850–0026 | 0.1666 | 3.91E-14 | 3.30E+35 | 6.75E+04 | 2.58E+12 | 92.9 | 15.6 |
| J1851–0029 | 0.5187 | 4.74E-15 | 1.30E+33 | 1.73E+06 | 1.59E+12 | 5.9 | 1.9 |
| J1852+0013 | 0.9578 | 1.40E-14 | 6.30E+32 | 1.08E+06 | 3.71E+12 | 4.0 | 1.5 |
| J1852+0305 | 1.3261 | 1.00E-16 | 1.70E+30 | 2.06E+08 | 3.72E+11 | 0.2 | 0.1 |
| J1853–0004 | 0.1014 | 5.57E-15 | 2.10E+35 | 2.88E+05 | 7.61E+11 | 74.0 | 12.3 |
| J1853+0505 | 0.9051 | 1.28E-15 | 6.80E+31 | 1.12E+07 | 1.09E+12 | 1.3 | 0.6 |
| J1853+0545 | 0.1264 | 6.12E-16 | 1.20E+34 | 3.27E+06 | 2.82E+11 | 17.7 | 4.0 |
| J1853+0853 | 3.9147 | 5.13E-15 | 3.40E+30 | 1.21E+07 | 4.54E+12 | 0.3 | 0.2 |
| J1855+0307 | 0.8453 | 1.81E-14 | 1.20E+33 | 7.40E+05 | 3.96E+12 | 5.5 | 1.9 |
| J1855+0422 | 1.6781 | 9.30E-16 | 7.80E+30 | 2.86E+07 | 1.26E+12 | 0.4 | 0.3 |
| J1855+0527 | 1.3935 | 2.67E-13 | 3.90E+33 | 8.26E+04 | 1.95E+13 | 10.0 | 3.2 |
| J1856+0102 | 0.6202 | 1.22E-15 | 2.00E+32 | 8.04E+06 | 8.81E+11 | 2.3 | 0.9 |
| J1856+0404 | 0.4203 | 3.70E-17 | 2.00E+31 | 1.80E+08 | 1.26E+11 | 0.7 | 0.3 |
| J1857+0526 | 0.3500 | 6.93E-15 | 6.40E+33 | 8.00E+05 | 1.58E+12 | 12.9 | 3.4 |
| J1858+0241 | 4.6932 | 2.43E-14 | 9.30E+30 | 3.06E+06 | 1.08E+13 | 0.5 | 0.3 |
| J1900–0051 | 0.3852 | 1.42E-16 | 9.80E+31 | 4.29E+07 | 2.37E+11 | 1.6 | 0.7 |
| J1901+0413 | 2.6631 | 1.32E-13 | 2.80E+32 | 3.21E+05 | 1.89E+13 | 2.7 | 1.2 |
| J1902+0723 | 0.4878 | 2.10E-16 | 7.10E+31 | 3.68E+07 | 3.24E+11 | 1.4 | 0.6 |
| J1904+0004 | 0.1395 | 1.18E-16 | 1.70E+33 | 1.87E+07 | 1.30E+11 | 6.7 | 1.9 |
| J1904+0738 | 0.2090 | 4.11E-16 | 1.80E+33 | 8.06E+06 | 2.97E+11 | 6.8 | 2.0 |
| J1905+0600 | 0.4412 | 1.11E-15 | 5.10E+32 | 6.28E+06 | 7.09E+11 | 3.6 | 1.3 |
| J1905+0616 | 0.9897 | 1.35E-13 | 5.50E+33 | 1.16E+05 | 1.17E+13 | 11.9 | 3.6 |
| J1907+0534 | 1.1384 | 3.15E-15 | 8.40E+31 | 5.73E+06 | 1.92E+12 | 1.5 | 0.7 |
| J1907+0740 | 0.5747 | 6.71E-16 | 1.40E+32 | 1.36E+07 | 6.29E+11 | 1.9 | 0.8 |
| J1907+0918 | 0.2261 | 9.43E-14 | 3.20E+35 | 3.80E+04 | 4.67E+12 | 91.3 | 15.8 |
| J1908+0457 | 0.8468 | 9.80E-16 | 6.40E+31 | 1.37E+07 | 9.22E+11 | 1.3 | 0.6 |
| J1908+0500 | 0.2910 | 2.59E-15 | 4.10E+33 | 1.78E+06 | 8.78E+11 | 10.4 | 2.8 |
| J1908+0839 | 0.1854 | 2.39E-15 | 1.50E+34 | 1.23E+06 | 6.73E+11 | 19.6 | 4.5 |
| J1910+0728 | 0.3254 | 8.31E-15 | 9.50E+33 | 6.21E+05 | 1.66E+12 | 15.7 | 4.0 |
| J1915+0752 | 2.0583 | 1.39E-16 | 6.30E+29 | 2.35E+08 | 5.41E+11 | 0.1 | 0.1 |
| J1916+0844 | 0.4400 | 2.90E-15 | 1.30E+33 | 2.40E+06 | 1.14E+12 | 5.9 | 1.9 |
| J1920+1040 | 2.2158 | 6.48E-15 | 2.40E+31 | 5.42E+06 | 3.83E+12 | 0.8 | 0.4 |
| J1920+1110 | 0.5099 | 1.56E-16 | 4.60E+31 | 5.18E+07 | 2.85E+11 | 1.1 | 0.5 |
| J1926+2016 | 0.2991 | 3.50E-15 | 5.20E+33 | 1.35E+06 | 1.04E+12 | 11.6 3.1 | |
| J1928+1923 | 0.8173 | 6.35E-15 | 4.60E+32 | 2.04E+06 | 2.31E+12 | 3.5 | 1.3 |

| Pulsar (B1950) | P (s) | $\dot{P}$ (s/s) | $\dot{E}$ (ergs/s) | $\tau$ (yrs) | $B_{surf}$ (G) | $B_{12}/P^2$ | 1/Q |
|---|---|---|---|---|---|---|---|
| J1931+1952 | 0.5011 | 1.01E-16 | 3.20E+31 | 7.88E+07 | 2.27E+11 | 0.9 | 0.4 |
| J1935+2025 | 0.0801 | 6.08E-14 | 4.70E+36 | 2.09E+04 | 2.23E+12 | 347 | 42 |
| J1941+2525 | 2.3062 | 1.61E-13 | 5.20E+32 | 2.27E+05 | 1.95E+13 | 3.7 | 1.5 |
| J1943+0609 | 0.4462 | 4.66E-16 | 2.07E+32 | 1.52E+07 | 4.61E+11 | 2.3 | 0.9 |
| J1946+2535 | 0.5152 | 5.64E-15 | 1.60E+33 | 1.45E+06 | 1.73E+12 | 6.5 | 2.1 |
| J2005+3547 | 0.6150 | 2.81E-16 | 4.80E+31 | 3.47E+07 | 4.21E+11 | 1.1 | 0.5 |
| J2005+3552 | 0.3079 | 2.99E-15 | 4.00E+33 | 1.63E+06 | 9.71E+11 | 10.2 | 2.8 |
| J2009+3326 | 1.4384 | 1.47E-15 | 1.90E+31 | 1.55E+07 | 1.47E+12 | 0.7 | 0.4 |
| J2011+3331 | 0.9317 | 1.79E-15 | 8.70E+31 | 8.27E+06 | 1.31E+12 | 1.5 | 0.7 |
| J2021+3651 | 0.1037 | 9.57E-14 | 3.40E+36 | 1.72E+04 | 3.19E+12 | 296 | 37 |
| J2030+3641 | 0.2001 | 6.50E-15 | 3.20E+34 | 4.88E+05 | 1.15E+12 | 28.7 | 6.2 |
| J0158+2106 | 0.5053 | 3.53E-16 | 1.08E+32 | 2.27E+07 | 4.27E+11 | 1.67 | 0.70 |
| J0229+2258 | 0.8068 | 2.81E-15 | 2.11E+32 | 4.55E+06 | 1.52E+12 | 2.34 | 0.96 |
| J0241+1604 | 1.5449 | 1.34E-15 | 1.44E+31 | 1.82E+07 | 1.46E+12 | 0.61 | 0.35 |
| J2139+00 | 0.3125 | — | — | — | — | — | — |

Notes: All values are taken from the ATNF Pulsar Catalog, apart from those of J0158+21, J0229+22 and J0241+16, which were provided by Kevin Stovall (private communication).

Table B2: Profile Geometry Information

| Pulsar | $P$ (s) | DM (pc/cm$^3$) | Class | $W_c$ (°) | $\alpha$ (°) | $R$ (°/°) | $\beta$ (°) | $W_i$ (°) | $\rho_i$ (°) | $\beta/\rho_i$ (°) | $W_o$ (°) | $\rho_o$ (°) | $\beta/\rho_o$ (°) |
|---|---|---|---|---|---|---|---|---|---|---|---|---|---|
| | | | (1-GHz Geometry) | | | | | (1.4-GHz Beam Sizes) | | | | | |
| J1802+0128 | 1.109 | 97.97 | Sd? | — | 53 | -12 | +3.8 | 3.5 | 4.1 | 0.94 | — | — | — |
| J1806+1023 | 0.484 | 52.03 | cQ/M? | — | 80 | ∞ | 0 | 13 | 6.3 | 0.00 | 16.8 | 8.3 | 0.00 |
| J1809–0119 | 0.745 | 140 | St/T? | 2.9 | **78** | ∞ | 0 | 10 | 5.0 | 0.00 | — | — | — |
| J1832+0029 | 0.534 | 28 | St? | 4.1 | **55** | ∞ | 0 | 14 | 5.7 | 0.00 | — | — | — |
| J1843–0000 | 0.880 | 101.5 | T | 4.5 | **35** | -6.4 | -5.2 | — | — | — | 11.2 | 6.0 | -0.86 |
| J1844+0036 | 0.461 | 345.54 | T | 11 | **20** | -3 | 6.5 | — | — | — | 28 | 8.4 | 0.77 |
| J1844–0030 | 0.641 | 605 | St? | 7.0 | **26** | -8 | 3.1 | — | — | — | — | — | — |
| J1845+0623 | 1.422 | 113 | D? | — | 63 | 18 | 2.8 | 5.0 | 3.6 | 0.78 | — | — | — |
| J1848–0023 | 0.538 | 30.6 | D | — | 60 | -11 | 4.5 | — | — | — | 14.4 | 7.8 | 0.58 |
| J1848+0604 | 2.219 | 242.7 | Sd | — | 49 | +24 | -1.8 | 6.1 | 2.9 | -0.62 | — | — | — |
| J1849+0127 | 0.542 | 207.3 | St? | 6.4 | **31** | +10 | 3.0 | 18 | 5.7 | 0.52 | — | — | — |
| J1850–0006 | 2.191 | 570 | Sd? | — | 10 | -6 | 1.7 | 24 | 2.8 | 0.59 | — | — | — |
| J1850–0026 | 0.167 | 947 | Scatt. | — | — | — | — | — | — | — | — | — | — |
| J1851–0029 | 0.519 | 510 | St/T? | 6.7 | **31** | 6 | 4.9 | 13 | 6.0 | 0.82 | — | — | — |
| J1852+0013 | 0.958 | 545 | St? | — | — | — | — | — | — | — | — | — | — |
| J1852+0305 | 1.326 | 320 | Sd? | — | 6 | 1.7 | 3.5 | 11.5 | 3.6 | 0.98 | — | — | — |
| J1853–0004 | 0.101 | 437.5 | ?? | — | — | — | — | — | — | — | — | — | — |
| J1853+0505 | 0.905 | 279 | Scatt. | — | — | — | — | — | — | — | — | — | — |
| J1853+0545 | 0.126 | 198.7 | Scatt. | — | — | — | — | — | — | — | — | — | — |
| J1853+0853 | 3.915 | 214 | Sd | — | 24 | +12 | 1.9 | 4.3 | 2.1 | 0.91 | — | — | — |
| J1855+0307 | 0.845 | 402.5 | T | 3 | **59** | +9 | 5.5 | — | 6.2 | 0.88 | 6.8 | 6.2 | 0.88 |
| J1855+0422 | 1.678 | 438 | D | — | 18.5 | +6 | 3.0 | 7.5 | 3.3 | 0.92 | — | — | — |
| J1855+0527 | 1.393 | 362 | Scatt. | — | — | — | — | — | — | — | — | — | — |
| J1856+0102 | 0.620 | 554 | cQ | — | 66 | 11 | 4.8 | 6 | 5.5 | 0.86 | 12 | 7.3 | 0.65 |
| J1856+0404 | 0.420 | 341.3 | Sd | — | 17 | +2 | 8.4 | — | — | — | 14 | 8.8 | 0.96 |
| J1857+0526 | 0.350 | 466.4 | Scatt. | — | — | — | — | — | — | — | — | — | — |
| J1858+0241 | 4.693 | 336 | Sd | — | 14.5 | +12 | 1.2 | 12.0 | 2.0 | 0.61 | — | — | — |
| J1900–0051 | 0.385 | 136.8 | Sd/D? | — | 36 | 7 | 4.8 | 16 | 6.9 | 0.70 | — | — | — |
| J1901+0413 | 2.663 | 352 | D | 0 | 18.4 | +11 | 1.6 | 12.8 | 2.7 | 0.62 | — | — | — |
| J1902+0723 | 0.488 | 105 | D/cT? | — | 45 | -8 | -5.1 | — | 8.2 | -0.62 | 19 | 8.2 | -0.62 |

| Pulsar | $P$ | DM | Class | $W_c$ | $\alpha$ | $R$ | $\beta$ | $W_i$ | $\rho_i$ | $\beta/\rho_i$ | $W_o$ | $\rho_o$ | $\beta/\rho_o$ |
|---|---|---|---|---|---|---|---|---|---|---|---|---|---|
| | (s) | (pc/cm$^3$) | | (°) | (°) | (°/°) | (°) | (°) | (°) | (°) | (°) | (°) | (°) |
| | | | | (1-GHz Geometry) | | | | (1.4-GHz Beam Sizes) | | | | | |
| J1904+0004 | 0.140 | 233.6 | ?? | — | — | — | — | — | — | — | — | — | — |
| J1904+0738 | 0.209 | 278.3 | St? | 4.9 | — | — | — | — | — | — | — | — | — |
| J1905+0600 | 0.441 | 730.1 | ?? | — | — | — | — | — | — | — | — | — | — |
| J1905+0616 | 0.990 | 256.1 | St | 3.1 | **53** | -10 | -4.6 | — | — | — | 9 | 5.7 | -0.80 |
| J1907+0534 | 1.138 | 524 | ?? | — | — | — | — | — | — | — | — | — | — |
| J1907+0740 | 0.575 | 332 | D | — | 39 | +7.5 | +4.8 | 9.8 | 5.8 | 0.83 | — | — | — |
| J1907+0918 | 0.226 | 357.9 | ?? | 3.0 | — | — | — | — | — | — | — | — | — |
| J1908+0457 | 0.847 | 360 | cQ? | — | 33 | -9 | 3.5 | 12 | 4.9 | 0.71 | 18.2 | 6.2 | 0.56 |
| J1908+0500 | 0.291 | 201.4 | T? | 6 | **53** | -5 | -9.2 | — | — | — | 14.5 | 10.6 | -0.86 |
| J1908+0839 | 0.185 | 512.1 | ?? | — | — | — | — | — | — | — | — | — | — |
| J1910+0728 | 0.325 | 283.7 | D | — | 37 | 4 | 8.6 | — | — | — | 15.6 | 10.0 | 0.86 |
| J1915+0752 | 2.058 | 105.3 | Sd | — | 40 | -14 | 2.6 | 4.4 | 3.0 | 0.88 | — | — | — |
| J1916+0844 | 0.440 | 339.4 | St | 4.2 | **62** | -6 | 8.4 | — | — | — | — | — | — |
| J1920+1040 | 2.216 | 304 | Sd | — | 13 | +4.8 | 2.7 | 7.6 | 2.9 | 0.94 | — | — | — |
| J1920+1110 | 0.510 | 182 | St/Sd? | 7.1 | **29** | -13 | 2.1 | 23 | 6.1 | 0.35 | — | — | — |
| J1926+2016 | | | | | | | | | | | | | |
| J1928+1923 | 0.817 | 476 | Scatt? | — | — | — | — | — | — | — | — | — | — |
| J1931+1952 | 0.501 | 441 | ?? | — | 90 | -24 | 2.4 | 5.8 | 3.8 | 0.64 | — | — | — |
| J1935+2025 | 0.080 | 182 | St? | 12.3 | **45** | +2.7 | 15.1 | — | — | — | — | — | — |
| J1941+2525 | 2.306 | 314.4 | St/T | 3 | **33** | +15 | +2.1 | 7.1 | 2.8 | 0.72 | — | — | — |
| J1943+0609 | | | | | | | | | | | | | |
| J1946+2535 | 0.515 | 248.8 | ?? | — | — | — | — | — | — | — | — | — | — |
| J2005+3547 | 0.615 | 401.6 | Scatt. | 0 | — | — | — | — | — | — | — | — | — |
| J2005+3552 | 0.308 | 455 | St? | 11.0 | **24** | — | — | — | — | — | — | — | — |
| J2009+3326 | 1.438 | 263.6 | D | — | 18 | +9 | +2.0 | 19.5 | 3.7 | 0.53 | — | — | — |
| J2011+3331 | 0.932 | 298.6 | T | 3.4 | **49** | -90 | 0.5 | 11.9 | 4.5 | 0.11 | — | — | — |
| J2021+3651 | 0.104 | 367.5 | Scatt. | — | — | — | — | — | — | — | — | — | — |
| J2030+3641 | 0.200 | 246 | ?? | — | — | — | — | — | — | — | — | — | — |
| J2139+00 | 0.312 | 31.2 | cT? | — | — | — | — | — | — | — | — | — | — |

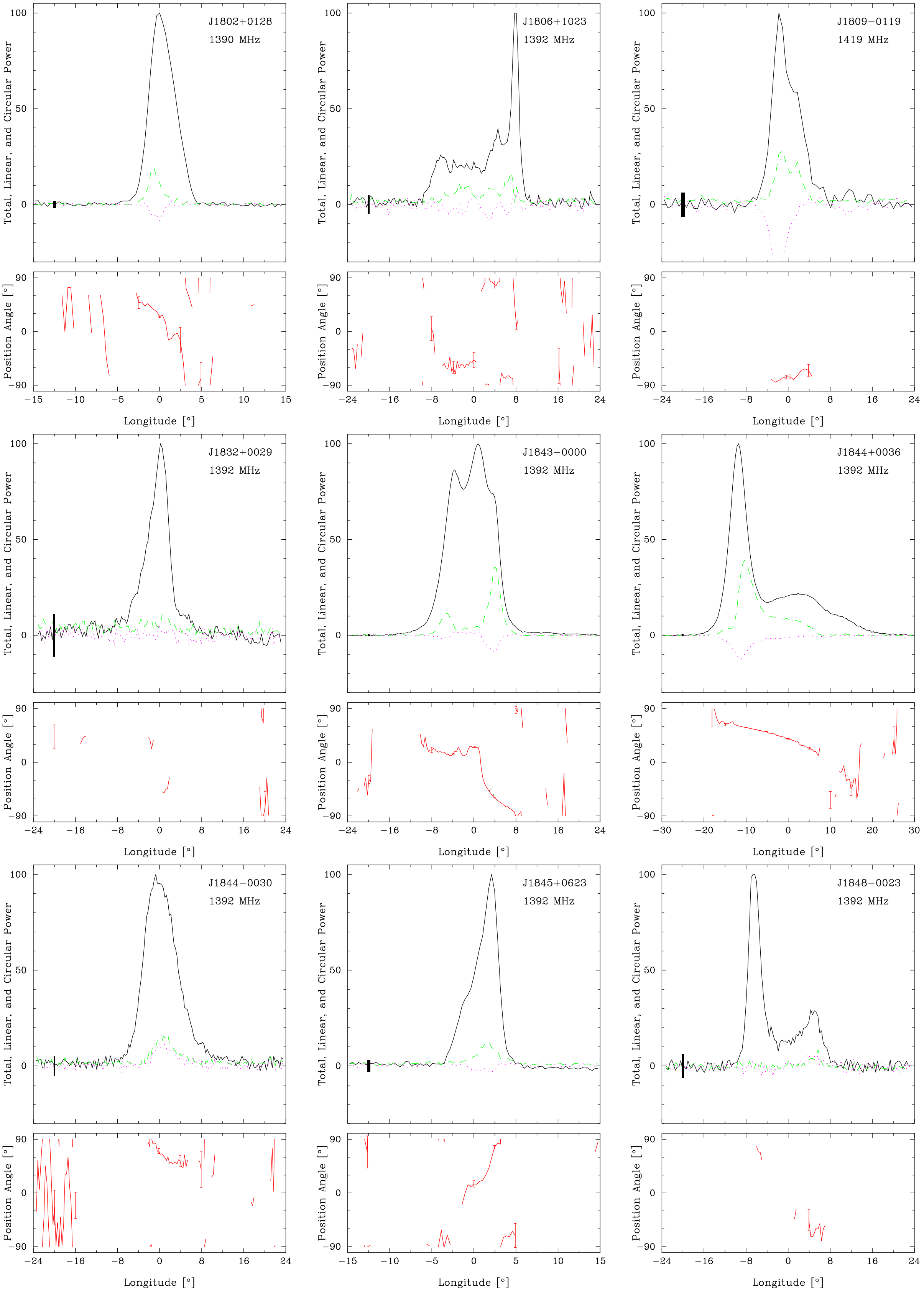

**Figure B1.** Average polarization profiles: the top panel in each profile gives the total intensity (solid black line), the linear polarization (dashed green) and circular polarization (dotted red)**—and a box to the left gives the resolution and the 3-$\sigma$ amplitude**. The bottom panels show the PPA against longitude **with 3-$\sigma$ error bars at intervals for values less than half a radian**.

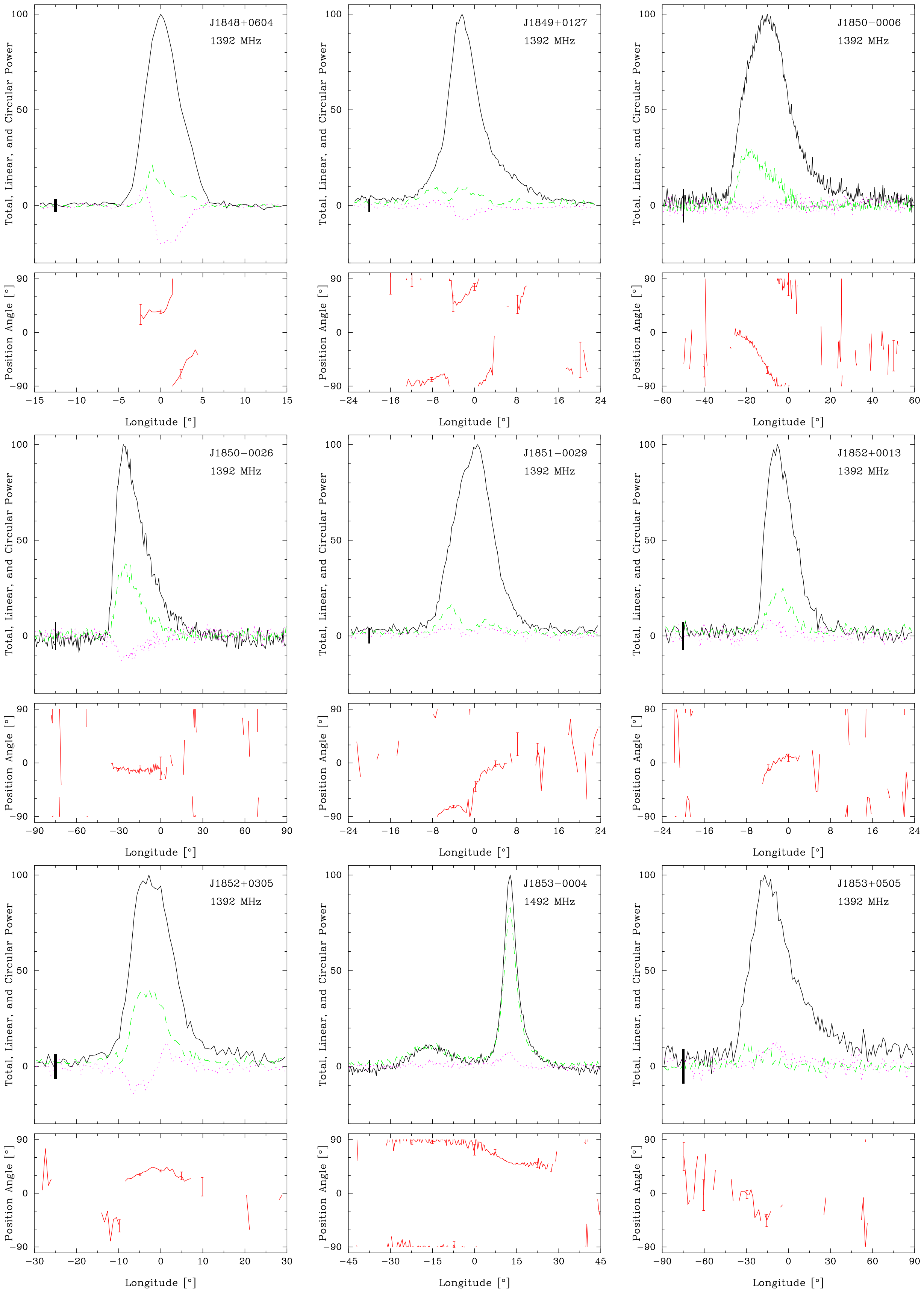

**Figure B2.** Average polarization profiles as in Fig. B1.

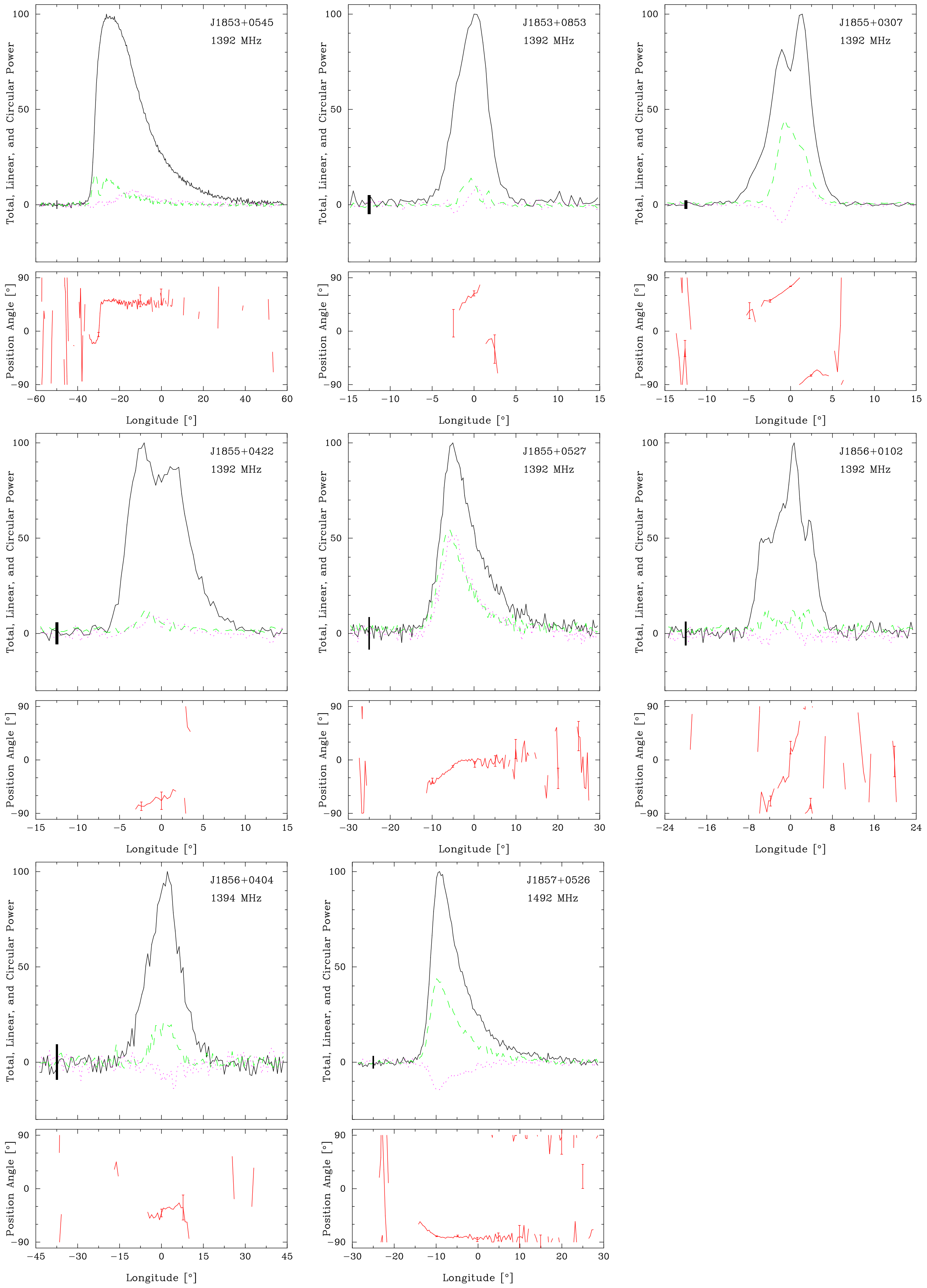

**Figure B3.** Average polarization profiles as in Fig. B1.

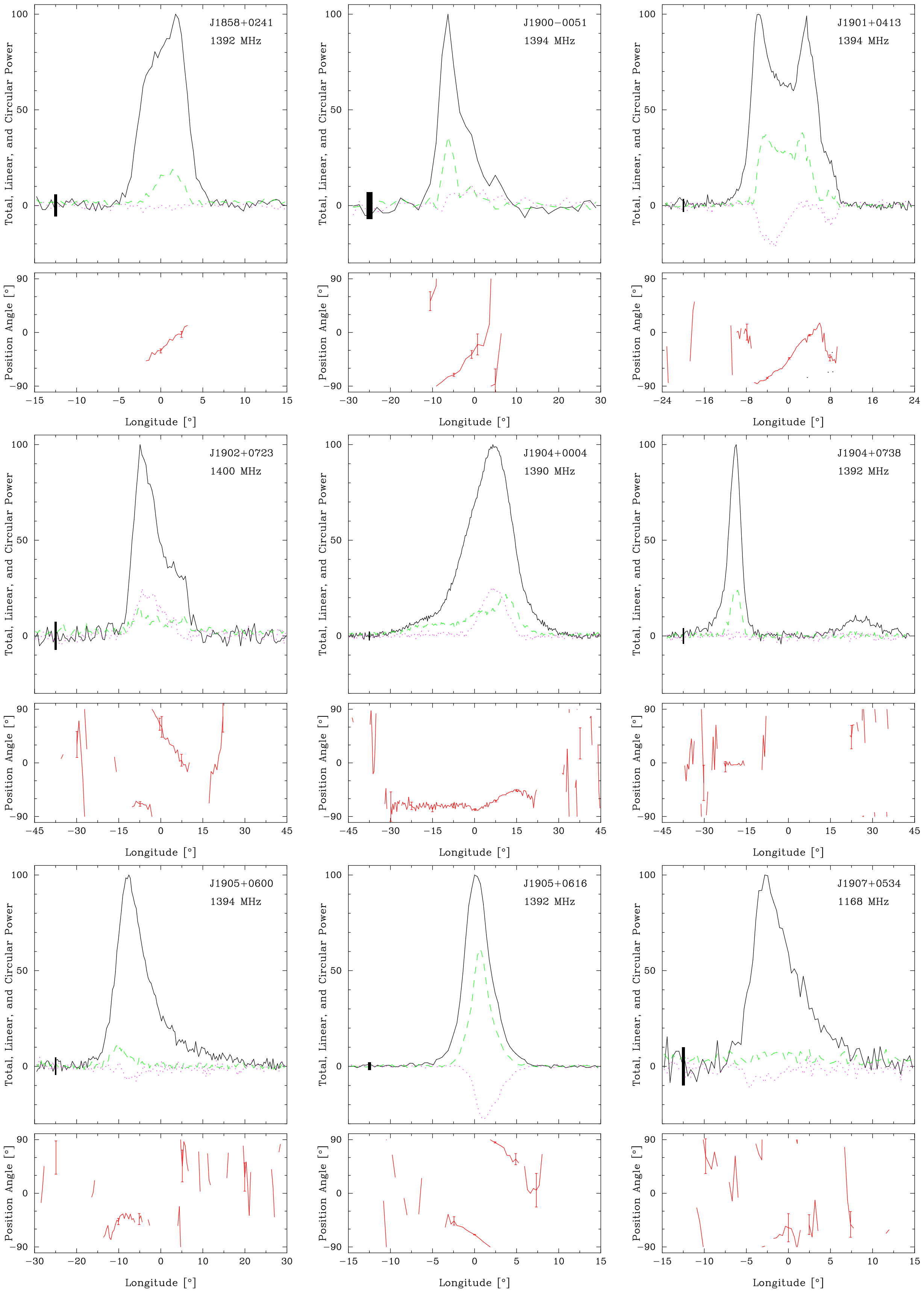

**Figure B4.** Average polarization profiles as in Fig. B1.

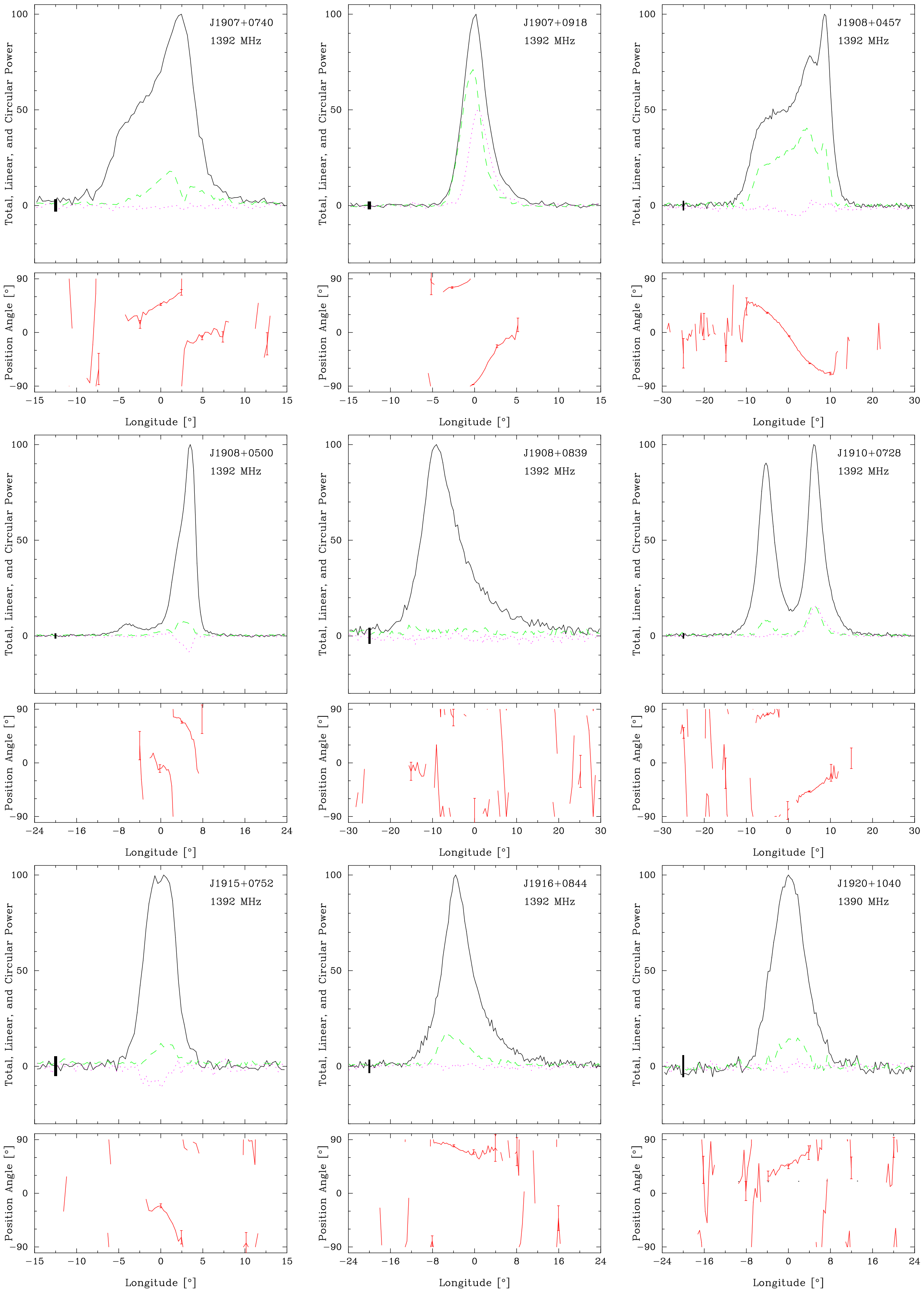

**Figure B5.** Average polarization profiles as in Fig. B1.

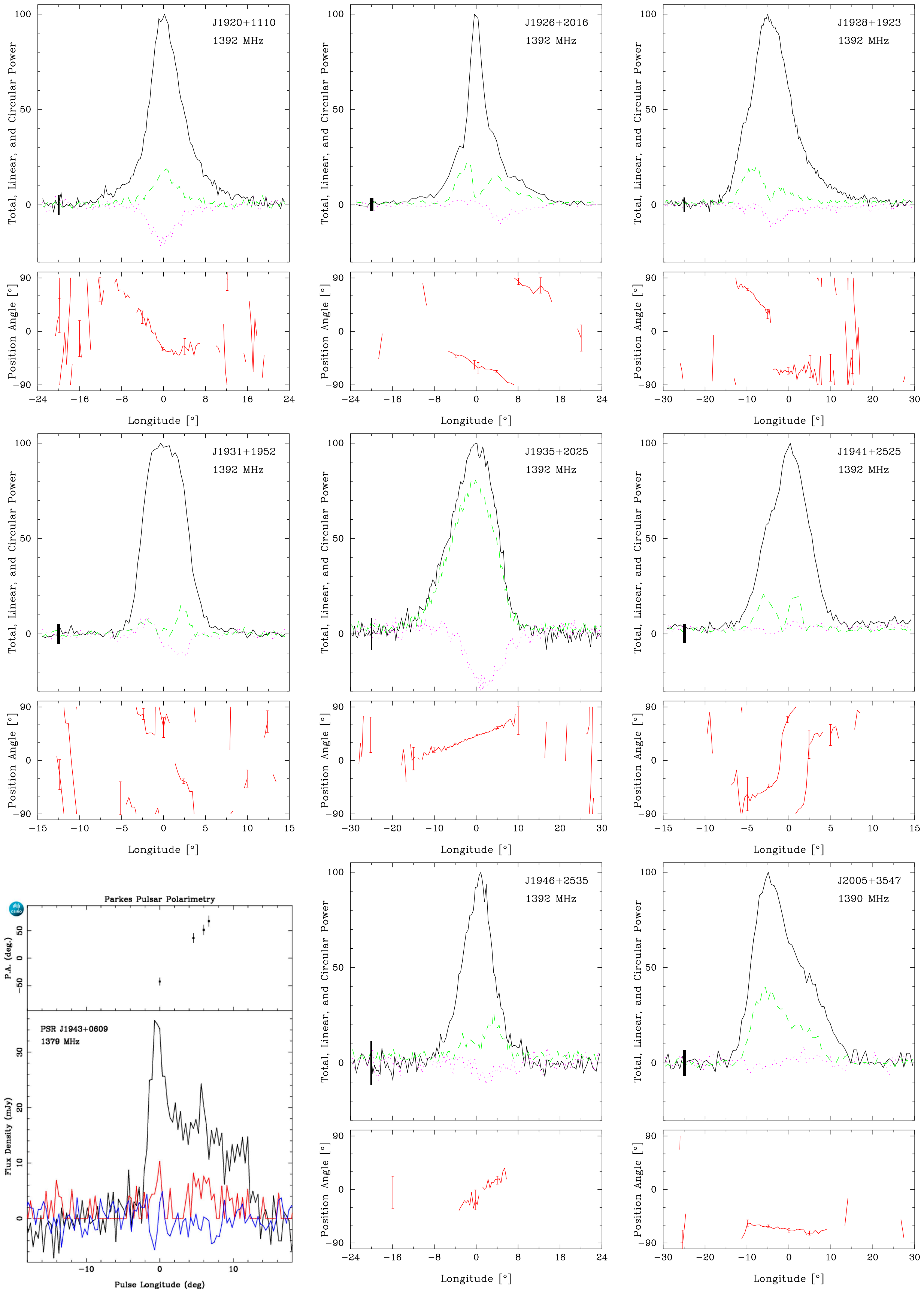

**Figure B6.** Average polarization profiles as in Fig. B1.

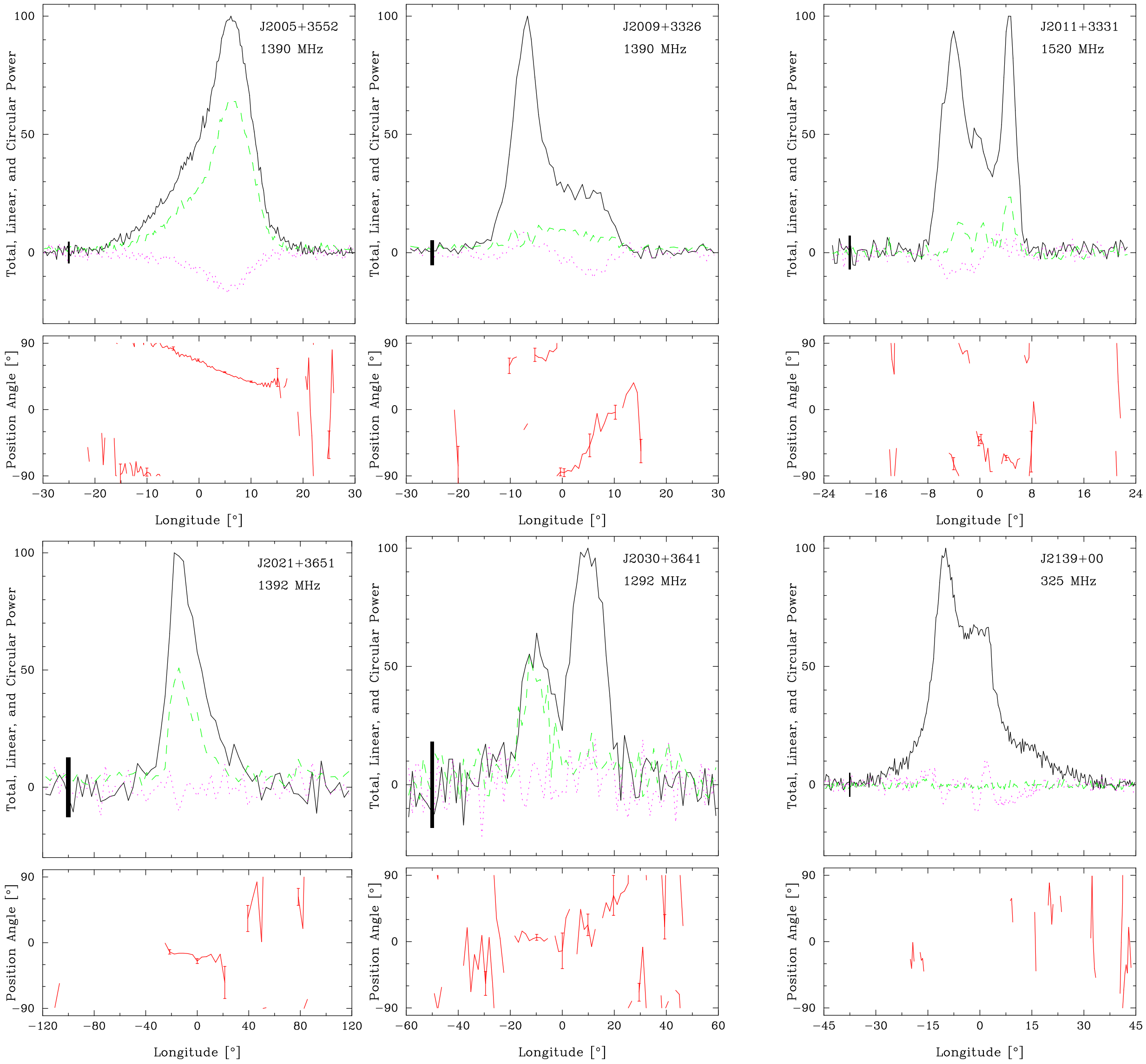

**Figure B7.** Average polarization profiles as in Fig. B1; J2139+00 is at 325 MHz.

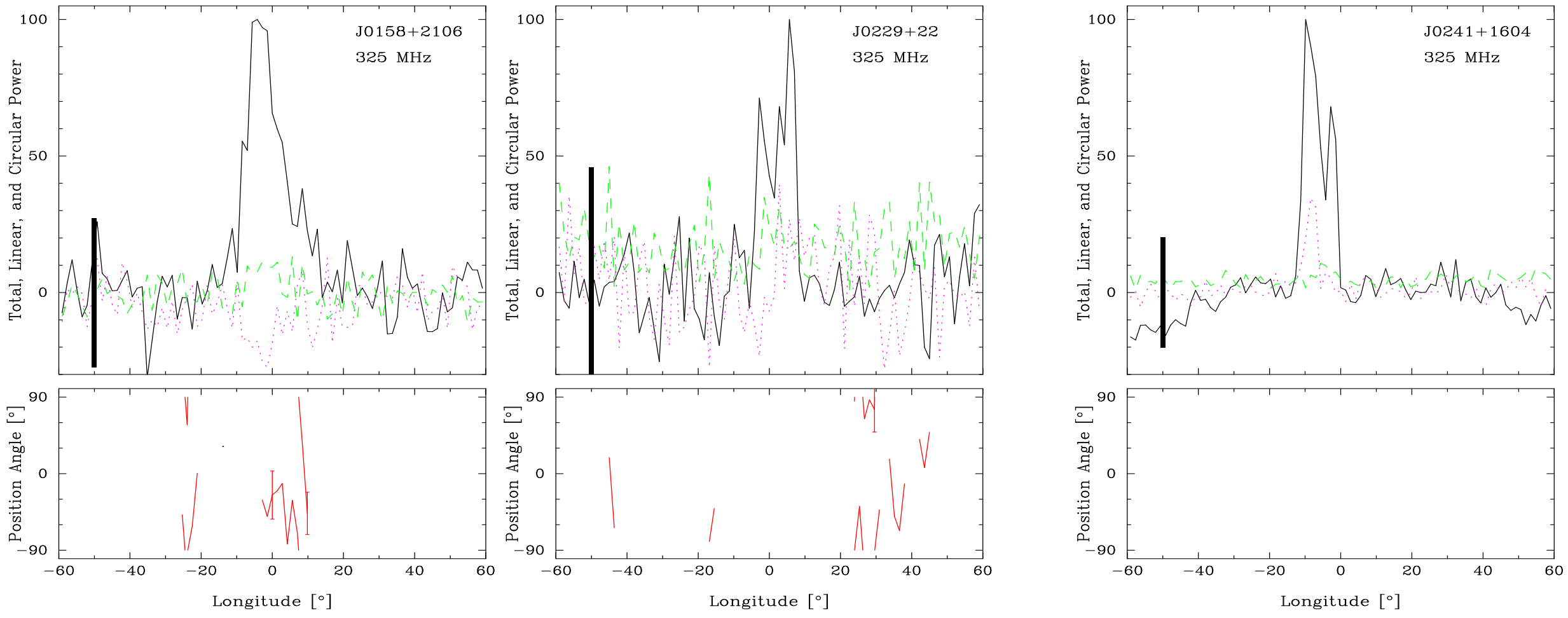

**Figure B8.** Average 325-MHz polarization profiles as in Fig. B1.

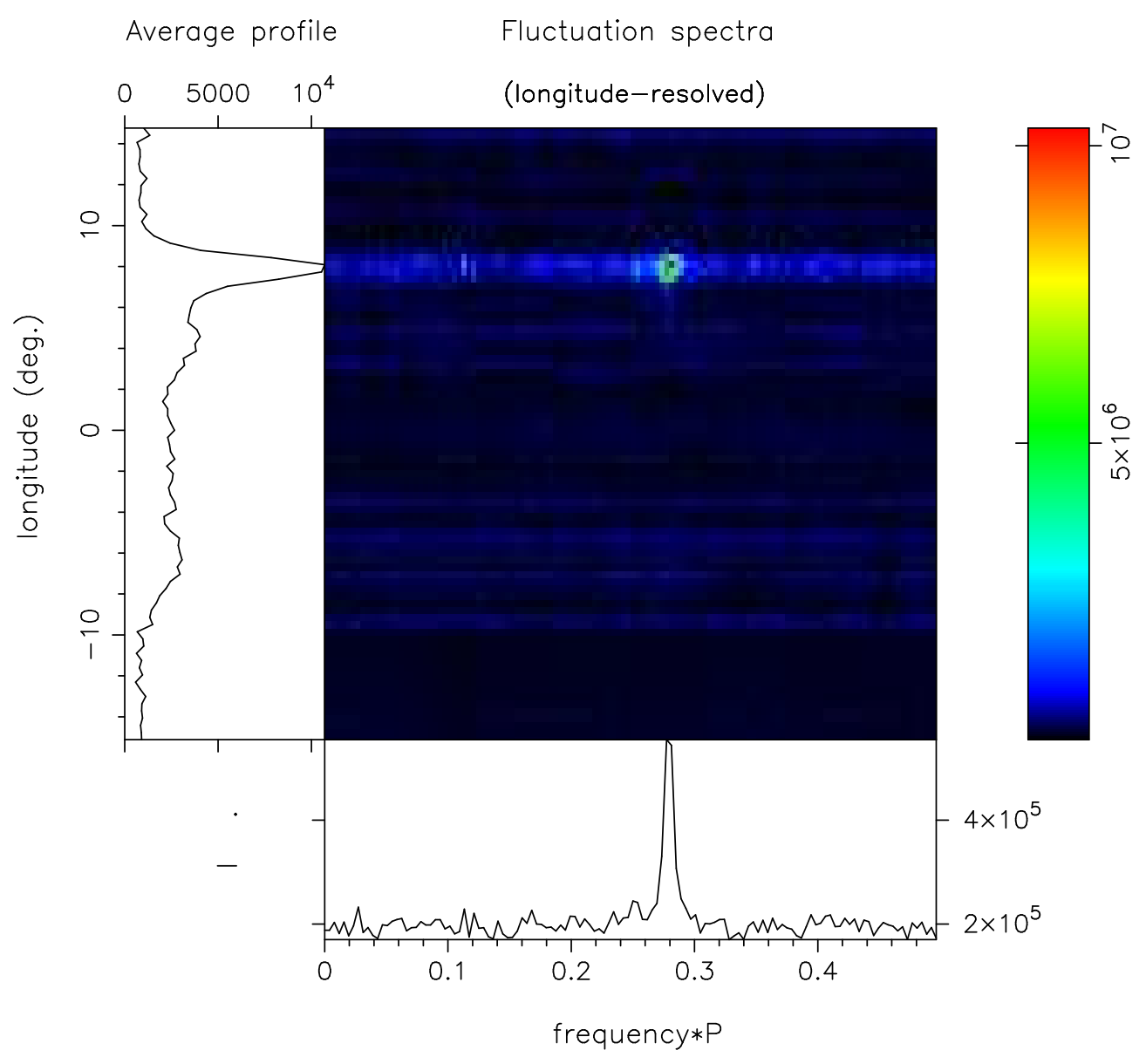

**Figure B9.** Pulsar J1806+1023 shows a strong 3-*P* fluctuation feature in its bright trailing conal component. A 256-length FFT was used.

***J1802+0128***, we find, is a 1-s, not a half-sec pulsar. Little is known of its frequency evolution, though McEwen et al. (2020) show a poor 350-MHz profile. We have guessed from the polarization and the above that it has a conal single configuration, though its featureless fluctuation spectra give no support to this end.

***J1806+1023*** has the "boxy" form that many five-component **M** profiles show at higher frequencies. We assume a central sightline traverse, and the inner cone dimension is estimated from the linear polarization.

***J1809–0119***: not much to go on here, but were the central feature a core, flanked by a strong leading and weak trailing conal outrider, the quantitative model roughly supports this interpretation. McEwen et al. (2020) do show a 350-MHz profile, but it fails to clarify the pulsar's evolution.

***J1832+0029*** is the Lorimer et al. (2012) intermittent pulsar. It appears to have a core-single $\mathbf{S}_t$ geometry, and if the weak pair of leading and trailing features are conal, they would have the right geometry to be the usual inner cone outriders.

***J1843–0000***: The three-component profile has a fluctuation spectrum with only "red" noise, so we model it with a core-cone triple **T** geometry. The McEwen et al. (2020) 350-MHz width value shows a broader profile with what may well be a scattering "tail". We model it with an outer conal geometry.

***J1844+0036*** shows a broad tripartite profile with a well defined negative PPA traverse. The widths can only be estimated, however, from the one profile. As the fluctuation spectrum is featureless, we model the profile as having a core and outer conal **T** configuration. This requires a core width of 10.7°, which seems quite plausible. Were it an inner cone, the core width would be more like 15°—and this cannot be ruled out. The McEwen et al. (2020) 350-MHz profile show substantial scattering.

The pulsar has no reported spin down in the ATNF catalogue. We initially noticed a small drift in the single pulses. Correcting this required a spin down of some 3.2x$10^{-15}$ s/s, and this value puts the pulsar well into the core-dominated regime as expected. We also note that the pulsar was well detected within Arecibo's 3.1' beam at 1.4 GHz, so the reported declination error must be smaller than the specified ±5 arc mins.

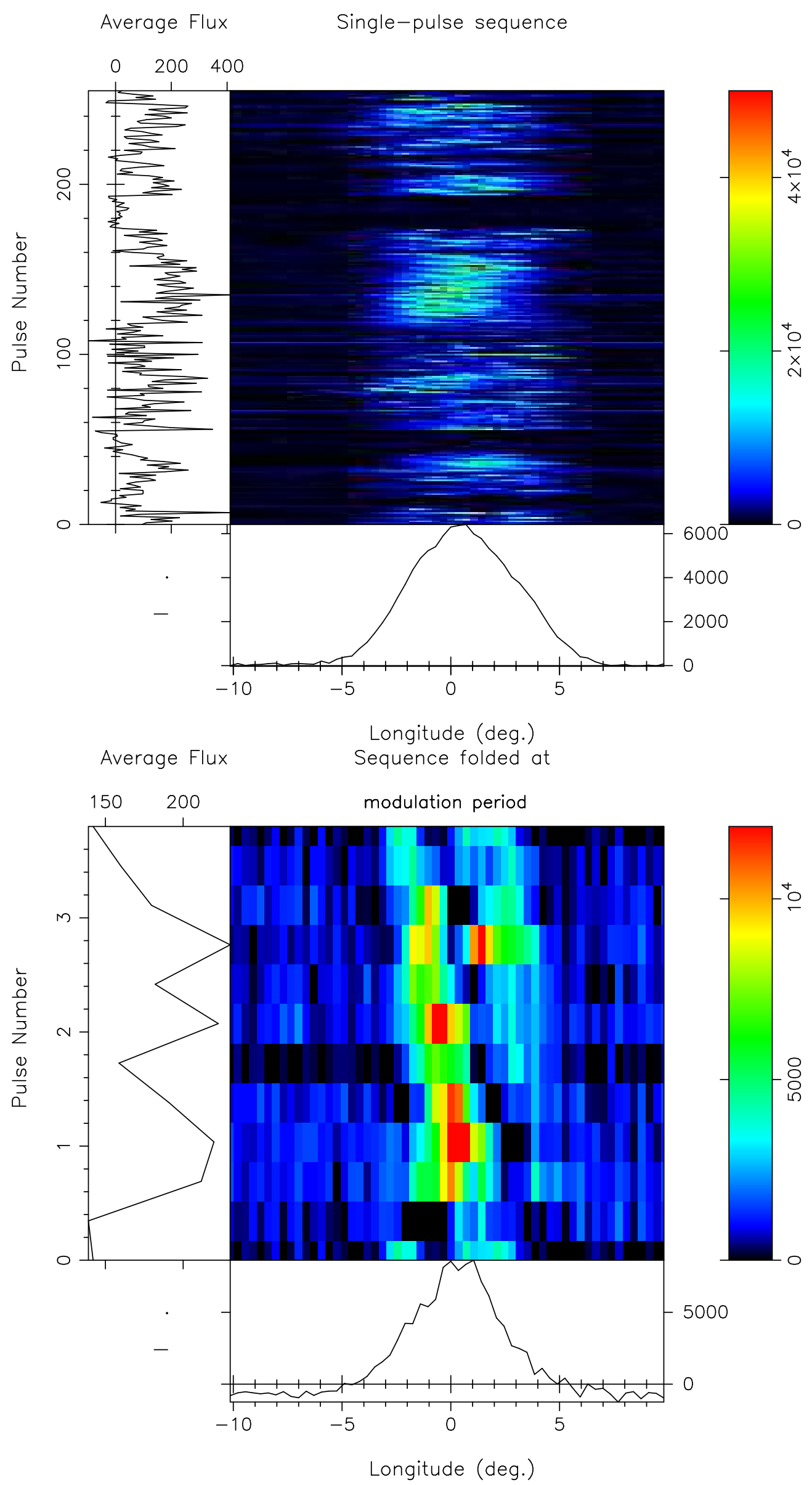

**Figure B10.** Pulsar J1842+0257 emits a complex mixture of drifting-like emission and long nulls (upper panel), and the pulses 120-170 show a clear 3.-period drift modulation.

***J1844–0030***: Probably a core-single pulsar, because we see a featureless fluctuation spectrum. However, we have not other way to confirm this.

***J1845+0623***: Despite that we see no sign of drifting in either the single pulses or fluctuation spectra, the profile very likely represents an inner conal barely resolved double **D** geometry, and so we have modeled it.

***J1848–0023*** has a wide double profile with little linear polarization; however, there is a hint of a negative PPA track. Its emission comes in short bursts with long intervals of nulls or weak pulses in between. Within the bursts one sees a hint of regular modulation with a period of a few pulses. We thus model it as having an outer conal **D** configuration. Unfortunately, McEwen et al. (2020) 350-MHz profile is too poor to be useful.

***J1848+0604*** shows substantial two-period fluctuation power, and the single pulses in bursts often have an odd-even modulation. Its geometry then seems most compatible with an inner conal single $\mathbf{S}_d$ one, especially given the break on the profile's trailing edge suggesting a second weak conflated feature—a frequent property of this type. Han et al. (2009) show a similar form and polarization at 774 MHz.

***J1849+0127*** emits in rough non-drifting bursts of about 5-7 periods. It may well be a core-single $\mathbf{S}_t$ profile similar to B0540+23. The PPA traverse is well defined, and if the "skirts" of the profile are incipient conal outriders with a width of about 18° their geometry is compatible with that of an inner cone.

***J1850–0006*** shows a broad profile with what may be a scattering tail—though the scattering cannot been too severe because the PPA traverse its still intact. The star emits with a strong roughly 23-period modulation. We model the star with an inner conal single geometry but without much conviction.

***J1850–0026***: We detect a broad profile with a flat PPA and a clear scattering tail. The fluctuation spectrum is flat. No modeling is possible, however it would be surprising if this star's emission is not core-dominated.

***J1851–0029***: The profile and polarization suggests a profile of several parts, and were the brightest part a core, its width can be estimated on its trailing side. Then if the conflated leading feature and weak trailing one are conal outriders, we see that their half-power width roughly squares with this interpretation. So our model represents a core-single or triple geometry.

***J1852+0013***: Probably a core-single profile with some scattering showing on its trailing edge. Its fluctuation spectrum is white, and the widths seem enough affected by scattering that no sensible modeling is possible.

***J1852+0305***: Probably a conal single $\mathbf{S}_d$ profile although the weak single pulses show no feature in the fluctuation spectra. The PPA rate is ambiguous, but adequate to estimate a geometrical model.

***J1853–0004***: The putative double profile of this 100-ms pulsar is fully linearly polarized but has a PPA traverse that cannot be interpreted within the single-vector model. We therefore can give no model.

***J1853+0505***: The profile is clearly scattered and the fluctuation spectrum give no clear information about the type of emission.

***J1853+0545***: The profile is clearly scattered and the fluctuation spectrum give no clear information about the type of emission.

***J1853+0853***'s fluctuation spectra show drift bands with irregular spacing of 5-7 periods. The pulsar then has a conal single $\mathbf{S}_d$ geometry, but whether an inner or outer cannot be determined. We model it as the former. Non-detection at 149 MHz per BKK+.

***J1855+0307***: The profile seems to have three parts, and a core width can be estimated from the central feature as well as a conal width from the weak leading and strong trailing feature. The quantitative geometry supports the core-cone triple **T** model as well as the outer cone.

***J1855+0422*** might have a conal double structure with some scattering seen on the trailing edge. We provisionally model it as such.

***J1855+0527***: The profile is clearly scattered and the fluctuation spectrum give no clear information about the type of emission.

***J1856+0102***'s unusual profile represents a sightline trajectory through both cones, and we model it as a conal quadruple c**Q** beam structure. The inner conal width can only be estimated at perhaps 6°, but is not critical to the modeling because the sightline traverse through it is so peripheral. Fluctuation spectra show a

***J1856+0404***: Almost certainly a conal single $\mathbf{S}_d$ beam, but no clear signature in the weak pulse sequence—and whether an inner or outer cone cannot be determined from this one observation.

***J1857+0526***: The profile is clearly scattered and the fluctuation spectrum give no clear information about the type of emission.

***J1858+0241***: The mission of this slow pulsar is comprised of short bursts of a few pulses with much longer intervals of nulls or weak pulses in between. We model it as a narrow inner conal double profile, but conal single may also be possible. This distinction has little effect of the quantitative geometry, but we note that an outer conal geometry is also very possible.

***J1900–0051***: The weak profile gives little to go on. Scattering seems unlikely, so it could have two or even three features. The PPA traverse is well defined, so we have modeled it using a conal double geometry, keeping in mind that many conal single profiles have an asymmetric strong leading/weak trailing form especially at higher frequencies.

***J1901+0413***: This pulsar emits very strong single subpulses at its trailing edge and perhaps other longitudes that may well be "giant" pulses. Our short 509-pulse observation has two of these, and this is the reason for the strange trailing-edge profile shape. The Johnston & Kerr (2018) 1.4-GHz profile shows a more symmetrical double structure. We then suppose the emission to be conal and model it as an inner one, but an outer cone is also possible with an $\alpha$ value of some 52°. If the core beam here is active, $\beta$ is well large enough to probably miss it.

***J1902+0723***: Little doubt that this pulsar's emission is conal as it has a fairly prominent 12-$P$ fluctuation feature. This said, the profile with a strong leading and weaker trailing components indicates a filled double structure, but nothing inside is resolved. Therefore, we model it with an outer conal double model, and indeed the 430-MHz profile of Camilo & Nice (1995) does seem to be broader. The PPA traverse is well defined and the structure is reminiscent of many conal triple profiles, so that further study may show inner conal emission as well.

***J1904+0004***: Opining meaningfully about this profile is difficult. McEwen et al. (2020) show a 350-MHz profile, but its width is almost entirely instrumental, and its asymmetry may indicate significant scattering. Other are available on the EPN website but none help to clarify the evolution. The flat early portion of the PPA traverse is strange and points the question about whether the profile shows one or two components. More study is needed here.

***J1904+0738***: Probably a core-single component with a postcursor very like the configuration of pulsar B0823+26. The PPA traverse is flat under the main pulse, and the putative core width is narrower than the polar cap size as is sometimes the case with cores at 21 cms. Nothing more can be done with a single frequency.

***J1905+0600***: The profile is clearly scattered and the fluctuation spectrum shows a 2.5-*P* feature suggesting conal modulation.

***J1905+0616*** seems to have a primary core beam, but its PPA traverse is unusually well defined if no conal emission is present. If the edges of the profile are so depolarized, then they have roughly the right width to be incipient conal outriders. Higher frequency profiles are needed to confirm this interpretation. The pulsar's emission comes in bursts of some 35 pulses with nulls or weak pulses in between.

***J1907+0534***: The profile is clearly scattered and the fluctuation spectrum gives no clear information about the type of emission.

***J1907+0740*** emits in isolated bursts every 25 pulses. It emission is almost certainly conal and we model it with a conal double **D** beam traverse. We use an inner cone, but it could be outer as there are no lower frequency profiles available to help resolve the issue.

***J1907+0918***: The strong positive *V* is confirmed by Johnston & Kerr (2018). It might be have a core-single profile configuration, but the profile width is narrower than that of the polar cap. Little more can be said without other frequencies to compare.

***J1908+0457***: The single pulses along with the profile strongly suggests that both cones are active in the emission, so we model the pulsar as an outer cone and then estimate what width the inner conal components would have to have. The value of about 12° seems very plausible to provide a satisfactory conal quadruple c**Q** beam model.

***J1908+0500***: Perhaps this is a core-cone triple configuration. The weak leading component is seen in the single pulses, and the well measured 430-MHz profile (Camilo & Nice, 1995) might have three features and an overall broader width. One must guess at the PPA rate but it might be some –5°/° with at least one 90° "jump". An outer conal beam requires and $\alpha$ value of 53°, and this in turn would imply a core width of just less that 6°, which seems quite plausible.

***J1908+0839***: The profile is clearly scattered and the fluctuation spectrum gives no clear information about the type of emission.

***J1910+0728***: We model this pulsar, of course as having an outer conal beam. No other profile seems available to study its evolution. For a star with so much rotational energy, we might expect to have prominent core emission, but the sightline impact angle $\beta$ is large enough in this configuration that a core beam would be missed.

***J1915+0752*** shows beautiful drifting subpulse bands with $P_3$ of 40-50 *P*, so definitely conal emission. The Han et al. (2009) 774-MHz profile has essentially the same width, so maybe the emission is inner conal. The PPA rate is steep and either an inner or outer conal beam is possible. We model it with an inner conal $\mathbf{S}_d$ geometry.

***J1916+0844***: Single pulses show what may be occasional "giant" pulses on the profile's trailing edge. The pulsar shows progressive scattering in our four bands. We model it as having a probable cone-single $\mathbf{S}_t$ geometry.

***J1920+1040***. Fluctuation spectra show bursts of drifting sub pulses (see Fig. B11, so there can be little doubt that the emission is conal. There seems to be no other profile at another frequency, so we can say nothing about the evolution. We thus model it as an inner cone, but it could have an outer conal geometry for a bit larger $\alpha$ value.

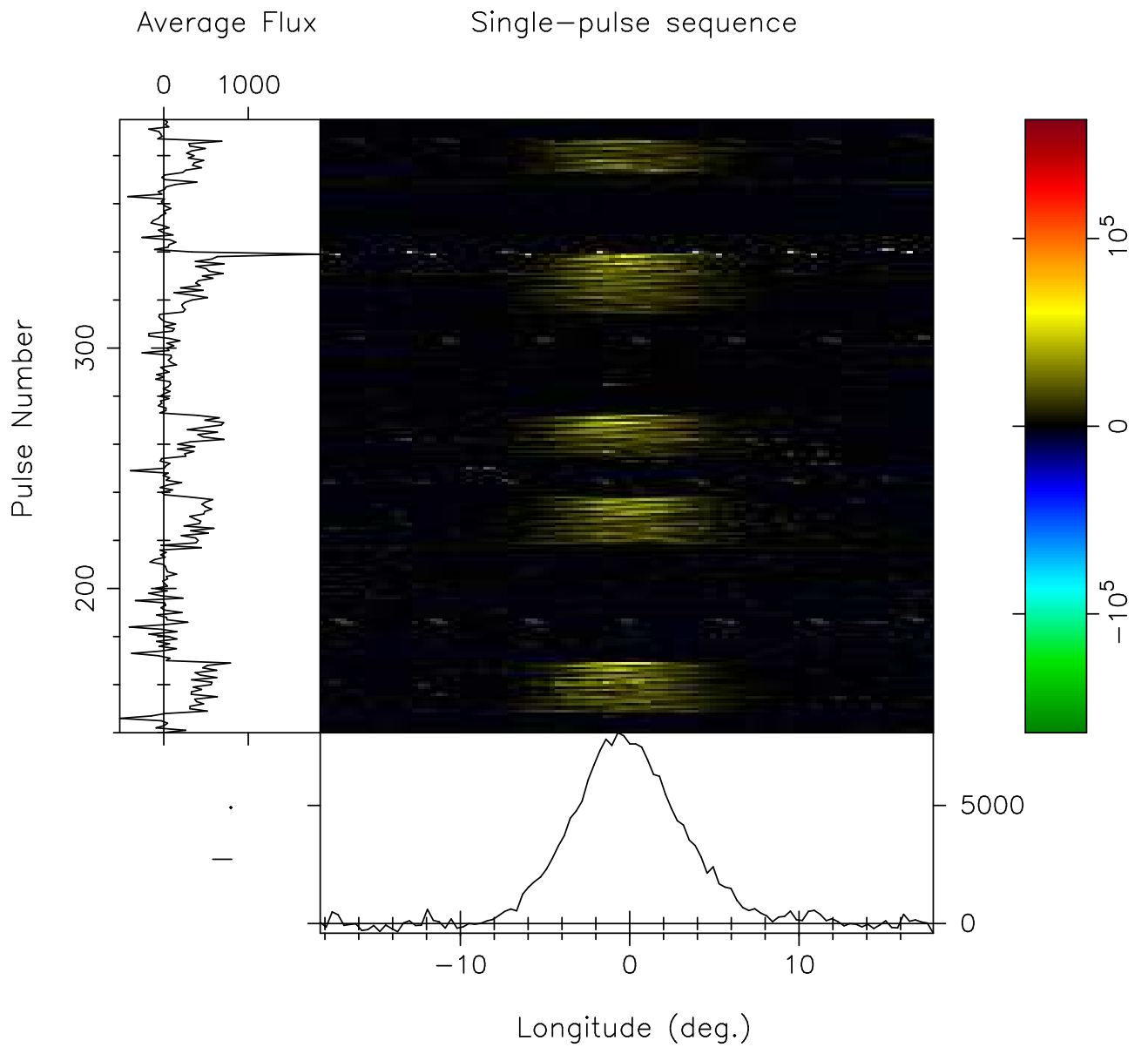

**Figure B11.** Pulsar J1920+1040 single pulses come in bursts of 20-30 pulses with deep nulls of usually longer duration in between. Within the bursts the sub pulses show a 2.5-*P* drift modulation. Radar RFI can be seen around pulse 339.

***J1920+1110***: The fluctuation spectra give no clarity about the pulsar's emission being conal. The PPA track is well defined, and the width is a bit too small to be an inner cone, even for an orthogonal geometry. Core single profiles do not usually have such an orderly PPA traverse, so perhaps a part of the "wing" emission is conal. A rough guess of this width is 23°, and this would represent an inner cone. This is the model we have used; however, there can be no surety until polarized observations are seen at other frequencies.

***J1926+2016***: The profile seems to have 3 or 4 features. If the bright central one is the core, it is too narrow to reflect the polarcap width. However, if the core includes the weaker feature on the trailing flank, an outer core model is possible with the well defined PPA rate and roughly 15°overall conal width.

***J1928+1923***: maybe this is a case of scattering just setting in. The trailing edge seems more affected than the leading with an apparent tail and flat PPA. No possibility of sensible modeling.

***J1931+1952***: It is unclear how to understand this pulsar. There are some fluctuation-spectral features suggesting the emission is conal, and the profile appears to have two conflated parts. However, the PPA rate is too steep for the narrow profile to be an inner cone. "More inner" cones have been suggested but never well confirmed. We show an orthogonal model in the table to exhibit the problem.

***J1935+2025***: Probably a core-single $\mathbf{S}_t$ pulsar, but little further can be said without having access to other frequencies and confirming the pulsar evolution.

***J1941+2525***: The intriguing PPA traverse as well as the linearly polarized emission suggest a three-part profile. It entails a +12°/° traverse interrupted by two 90° "jumps" that coincide with the depolarized points in the linear profile. Fluctuation spectra show no periodicities. We estimate the conal width using the break points and the linear profile, and model it with a core-cone triple **T** structure. if the central core opponent has a 3° width, it is an inner cone, if about 2°, an outer one.

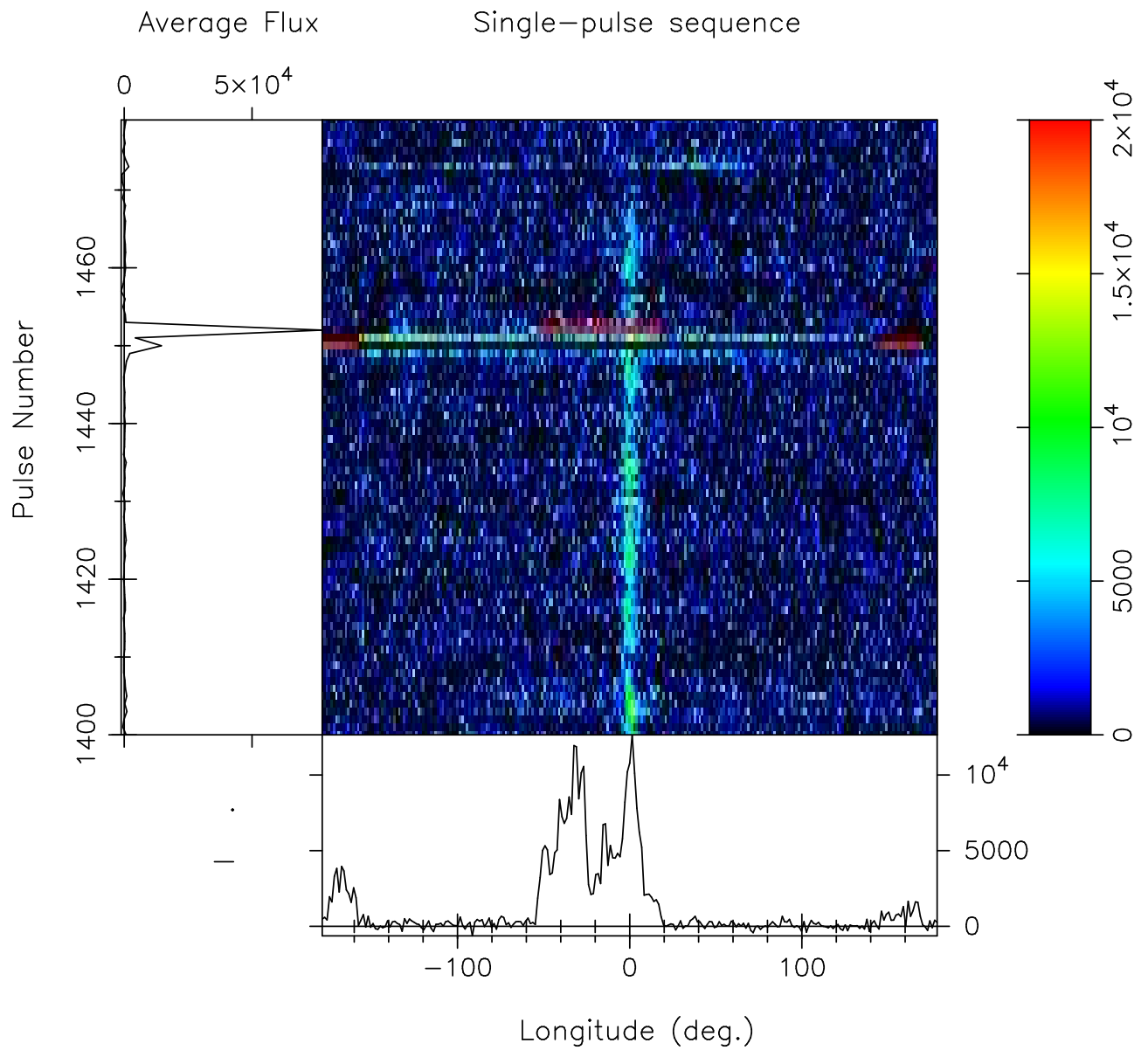

**Figure B12.** Pulsar J1920+1110: FLARE or RFI? We see the pulsar emit steadily up to about pulse 1470 and then tail off into a long null for the remainder of the 1946-pulse observation. Just before null interval, however, (pulse 1450) we see bright emission around the (unseen) interpulse longitude, and two pulses later (1452) bright emission before and at the main pulse longitude. These two events are the only bright emission in the observation.

***J1943+0609***: The Johnston (2021) 1.4-GHz observation (shown) suggests the "boxy" form commonly seen in five-component profiles at higher frequencies Unshown are 692-MHz and 3.1-GHz profiles, and none of them define the PPA rate. A core width might be about 4°, suggesting a nearly orthogonal geometry. If so—dealing only with the overall putative outer cone—the PPA rate has to be about 9°/° to give the needed radius, but this seems implausibly shallow. Were the rate much steeper as is usual in such profiles, the radius is about right for an inner cone—but what of the interior pair of components? Higher quality profiles are needed to resolve this pulsar's geometry.

***J1946+2535***: Fluctuation spectra give no information about this brief observation. Our conjecture is that it is a core beam, but further study at other frequencies is needed to do any modeling.

***J2005+3547***: The profile is clearly scattered and the single pulses show irregular intervals of emission together with short nulls. The fluctuation spectrum shows weak features. The emission is thus almost surely conal.

***J2005+3552*** has an unusual "reverse scattered" profile, and its emission comes in bursts of 10-20 pulses as shown by rough features in its fluctuation spectra. We cannot model it confidently with the core-double-cone model.

***J2009+3326***: The fluctuation spectra seem to show some conal features. We model the profile with an inner cone beam and a conal double **D** geometry, though an outer cone is also possible with a little larger $\alpha$ value.

***J2011+3331*** gives no hints in the fluctuation spectra. The steep PPA traverse and three-component structure suggests another core-cone **T** beam system. Again, a steep PPA traverse is suggested as well as a narrow central feature. If the central feature is a core, and its width a plausible 3.4°, then an inner cone model describes the geometry.

***J2021+3651***: The profile shows an exponential tail and a flat PPA, so it is surely scattered. In fact, we see the tail develop in our four bands from little of nothing at 1620 MHz to pronounced at 1170 MHz. The pulsar is very powerful in terms of spin down energy, and we expect its emission will be core dominated, but no higher frequency observations are available to study its evolution.

***J2030+3641*** has a strange double profile with leading component highly linearly polarized and the trailing one mostly depolarized. Moreover, the PPA under the first component is flat and the second possibly with a positive ramp. Maybe some clarity would come from seeing profiles at other frequencies, but without other information there is little to say. There are a very few other pulsars with profile showing such effects. One possibility is that the highly polarized leading feature is a "precursor" but this would need to be established.

***J2139+00***: The pulsar seems to have a broad three-part profile. However, there is no discernible linearly polarized power so the PPA traverse gives us nothing to go on. A guess is that it has a conal triple geometry, but other evidence such as from a fluctuation spectrum is needed to clarify this.